\documentclass[useAMS,usenatbib]{mn2e}
\usepackage{graphicx}
\usepackage[section]{placeins}
\DeclareGraphicsExtensions{.pdf,.png,.jpg,.eps}
\usepackage{epstopdf}
\usepackage{color}
\usepackage{amsmath}
\usepackage{amssymb}
\usepackage{amsbsy}
\usepackage{comment}
\definecolor{red}{rgb}{0,0,0}
\definecolor{blue}{rgb}{0,0,0}

\newcommand{\green}[1]{\iffalse #1 \fi}

\title[Performance of internal Covariance Estimators for Cosmic Shear Correlation Functions]{Performance of internal Covariance Estimators for Cosmic Shear Correlation Functions}
\author[O. Friedrich, S. Seitz, T. F. Eifler, D. Gruen]{\parbox{\textwidth}{O. Friedrich$^{1,2}$\thanks{E-mail:
oliverf@usm.uni-muenchen.de}, S. Seitz$^{1,2}$, T. F. Eifler$^{3,4}$, D. Gruen$^{1,2}$}\vspace{0.4cm}\\ \parbox{\textwidth}{$^{1}$University Observatory Munich,
Scheinerstrasse 1, 81679 Munich, Germany\\
$^{2}$Max Planck Institute for Extraterrestrial Physics,
Giessenbachstrasse, 85748 Garching, Germany \\
$^{3}$Jet Propulsion Laboratory, California Institute of Technology, 4800 Oak Grove Dr., Pasadena, CA 91109\\
$^{4}$California Institute of Technology, Pasadena, CA 91125, USA \\
}}

\begin{document}

\maketitle

\label{firstpage}

\begin{abstract}

\textcolor{blue}{\noindent Data re-sampling methods such as delete-one jackknife, bootstrap or the sub-sample covariance are a common tool for estimating the covariance of large scale structure probes. We investigate different implementations of these methods in the context of cosmic shear two-point statistics. Using log-normal simulations of the convergence field and the corresponding shear field we carry out realistic tests of internal covariance estimators. For a survey of $\sim 5000 \deg^2$ we find that jackknife, if implemented in what we call the \emph{galaxy-scheme}, provides the most reliable covariance estimates. Bootstrap, in the common implementation of duplicating sub-regions of galaxies, strongly overestimates the statistical uncertainties. }

In a forecast for the complete 5-year DES survey we show that internally estimated covariance matrices can provide a large fraction of the true uncertainties on cosmological parameters in a 2D cosmic shear analysis. The volume inside contours of constant likelihood in the $\Omega_m$-$\sigma_8$ plane as measured with internally estimated covariance matrices is on average $\gtrsim 85\%$ of the volume derived from the true covariance matrix. The uncertainty on the parameter combination $\Sigma_8 \sim \sigma_8 \Omega_m^{0.5}$ derived from internally estimated covariances is $\sim 90\%$ of the true uncertainty.
\end{abstract}


\section{Introduction}

Two-point statistics of cosmological random fields such as the cosmic shear correlation functions or the galaxy clustering angular correlation function are common probes of the large scale structure of the universe. Recent measurements of these correlation functions are e.g. reported in \citet{Thomas2011, Kilbinger1, deSimoni2013, Becker2015}. In order to use these statistics for constraining cosmological models one needs a quantitative description of the joint distribution of the correlation function estimators. When assuming multivariate Gaussian errors, this is given by the covariance matrix. On large angular scales this covariance matrix can - both for cosmic shear and galaxy clustering - be well described by a Gaussian approximation for the involved fields \citep{Schneider2002, Crocce2011}. It has, however, been shown, that the Gaussian approximation fails to describe the true PDF of the weak lensing convergence field \citep{Taruya2002,Vale2003} and that it underestimates the true covariance of the cosmic shear correlation functions on small scales, which can be alleviated by an empirical re-scaling \citep{Semboloni,Sato2011}, a log-normal approximation \citep{Hilbert2011}, or halo model approaches \citep[e.g.][]{Cooray,Takada,Eifler2014}.

Alternatives to modelling the covariance matrix are to estimate it from many independent realisations of cosmological N-body simulations or to estimate it internally, i.e. from the data itself. The latter method is independent of assuming a particular cosmological model and is hence often used to complement the other methods (\citealt{Kilbinger1, Wang2013, Becker2015}.). 

So far the performance of internal covariance estimators has only been systematically studied for the galaxy clustering 2-pt function (in most detail by \citealt{Norberg2009}) or for cross-correlations of the Cosmic Microwave Background (CMB) and the galaxy field \citep{Cabre2007}. In our paper, we will concentrate on cosmic shear correlation functions. We will show that the shape noise part of the covariance can be very accurately estimated internally while the cosmic variance part is generally underestimated. Gaussian simulations of the convergence field hence yield an overly optimistic test of internal covariance estimators, since the Gaussian model underpredicts the cosmic variance contribution to the covariance. We overcome this problem by employing log-normal simulations of the convergence field.

In our paper we want to study the performance of internal covariance estimators such as bootstrap, jackknife or the sub-sample covariance. There is no complete agreement in the literature yet on whether internal covariance estimates can be used to constrain cosmological parameters from measured 2pt-correlations or whether they are a mere tool to generate reasonable errorbars in plots of correlation functions (see e.g. \citealt{Norberg2009,Wang2013, deSimoni2013,Taylor2012}). We want to address the questions of how many internal resamplings are required in order to get a stable covariance matrix, whether internal estimators over- or underestimate the covariance matrix and whether/how internal covariance estimates can yield unbiased estimates of the \emph{inverse} covariance matrix.

Our paper is organized as follows: In section \ref{sec:cosmic_shear_basics} we introduce the cosmic shear correlation functions and explain the Gaussian and the log-normal model for the covariance of 2-pt. function estimators. In section \ref{sec:log-normal_simulations} we describe the simulations we use to generate mock shape catalogues that follow any given input power spectrum and whose underlying convergence field has a log-normal PDF. These are the simulations with which we will test the performance of internal covariance estimators.

In section \ref{sec:jackknife_schemes} we introduce two distinct ways of performing jackknife estimation of the covariance of two-point measures - the pair-jackknife and the galaxy-jackknife. Furthermore, we are explaining why jackknife, bootstrap and subsample covariance are almost equivalent.

In section \ref{sec:jackknife_results} we apply internal covariance estimators to simulated cosmic shear surveys. We show that in the pair-scheme all estimators are almost identical and we demonstrate the systematic effects of the different estimation schemes when varying the number of re-samplings. Our method to find optimal estimation schemes has to be re-run for any specific survey, because the performance of internal estimators depends crucially on the depth and area of a survey. In the end of section \ref{sec:jackknife_results} we configure our simulations to match the complete, 5-year Dark Energy Survey (DES, \citealt{DES2005, Flaugher2005}) and test the accuracy of jackknife covariance matrices for this particular setting. The code used for our simulations is made publicly available\footnote{www.usm.uni-muenchen.de/people/oliverf/, the code also contains many other useful features, that e.g. enable the user to create mock data suitable for galaxy-galaxy lensing or galaxy clustering measurements.}.

In section \ref{sec:conclusions} we discuss the results of our work.

\section[]{Cosmic Shear Basics}
\label{sec:cosmic_shear_basics}

\subsection{Cosmic Shear Correlation Functions}

Cosmic shear measures the correlated distortion of galaxy shapes due to gravitational lensing by the large scale structure of the universe as a function of the angular distance of galaxy pairs on the sky. We follow here the notation of \citet{Schneider2002} and employ the flat-sky-approximation, i.e. we assume a tangential Cartesian coordinate system $\boldsymbol{\vartheta} = (\vartheta_1, \vartheta_2)$ on the sky. 

In this coordinate system the comic shear field is at each point characterized by a complex number $\boldsymbol \gamma(\boldsymbol{\vartheta}) = \gamma_1 + i\gamma_2$. If the separation vector $\Delta\boldsymbol{\vartheta} = \boldsymbol{\vartheta}_2 -\boldsymbol{\vartheta}_1$ of two points on the sky has the polar angle $\phi$ then the \emph{tangential} and \emph{cross} components of $\boldsymbol \gamma$ at $\boldsymbol{\vartheta}_2$ and $\boldsymbol{\vartheta}_1$ (with respect to each other) are defined as
\begin{equation}
\gamma_t = -\mathcal{R}\mathrm e \left(\boldsymbol \gamma e^{-2i\phi}\right) \ \ \ \ \ \ ;\ \ \ \ \ \ \gamma_\times = -\mathcal{I}\mathrm m \left(\boldsymbol \gamma e^{-2i\phi}\right)\ .
\end{equation}
The \emph{cosmic shear correlation functions} $\xi_\pm(\theta)$ are defined as the expectation values
\begin{equation}
\xi_\pm(\theta) = \langle \gamma_{t,1} \gamma_{t,2} \rangle \pm \langle \gamma_{\times,1} \gamma_{\times,2} \rangle\ ,
\end{equation}
where $\theta$ is the absolute value of $\Delta\boldsymbol{\vartheta}$. It can be computed in terms of the power spectrum $P_\kappa(\ell)$ of the scalar \emph{convergence field} $\kappa(\boldsymbol \vartheta)$ as
\begin{equation}
\label{eq:xi_to_P}
\xi_\pm(\theta) = \frac{1}{2\pi} \int \mathrm d \ell\ \ell\ P_\kappa(\ell) J_{0/4}(\ell\theta)\ ,
\end{equation}
where $J_{0}(x)$ ($J_4(x)$) is the $0$-th order ($4$-th order) Bessel function.

The shape of a galaxy can be characterized by a complex number $\boldsymbol{\epsilon}$ which is to first order the sum of the intrinsic shape $\boldsymbol{\epsilon}^{\mathrm{in}}$ of the galaxy and the distortion caused by gravitational lensing, i.e. the value $\boldsymbol \gamma(\boldsymbol \vartheta)$ at the location $\boldsymbol \vartheta$ of the galaxy,
\begin{equation}
\boldsymbol{\epsilon} = \boldsymbol{\epsilon}^{\mathrm{in}} + \boldsymbol \gamma\ .
\end{equation}

In a cosmic shear survey the shapes $\boldsymbol{\epsilon}_i$ of many galaxies are measured and (cf. \citealt{Schneider2002}) an estimator for the correlation function can be constructed as
\begin{equation}
\label{eq:xi_pm_estimator}
\hat \xi_\pm(\theta) = \frac{\sum_{ij} w_j w_j (\epsilon_{t,i}\epsilon_{t,j}\pm\epsilon_{\times,i}\epsilon_{\times,j}) \Delta_\theta(ij)}{\sum_{ij} w_j w_j \Delta_\theta(ij)}\ ,
\end{equation}
where we have allowed for some weighting scheme $w_i$ for the shape measurements and where the filter $\Delta_\theta(ij)$ selects all galaxy pairs $(i,j)$ in the survey whose angular separation lies in some finite bin around $\theta$. The normalization in equation \ref{eq:xi_pm_estimator} is the effective number of galaxy pairs in a bin around $\theta$, which we will abbreviate as
\begin{equation}
N_p(\theta) = \sum_{ij} w_j w_j \Delta_\theta(ij)\ .
\end{equation}

\subsection{Covariance of the Correlation Functions}
\label{sec:covariance_models}

The covariance matrix of the estimator in equation \ref{eq:xi_pm_estimator} is defined as
\begin{eqnarray}
\label{eq:covariance_definition}
C_{\pm, \pm}(\theta_1, \theta_2) &=& \langle(\hat \xi_\pm(\theta_1) - \xi_\pm(\theta_1))(\hat \xi_\pm(\theta_2) - \xi_\pm(\theta_2))\rangle \nonumber \\
&=& \langle\hat \xi_\pm(\theta_1)\hat \xi_\pm(\theta_2)\rangle  - \xi_\pm(\theta_1)\xi_\pm(\theta_2)\ .
\end{eqnarray}
In order to compute this covariance matrix it is convenient to split $\xi_\pm(\theta)$ into the three different contribution
 \begin{eqnarray}
 \label{eq:xi_contributions}
 \hat \xi_\pm^{nn}(\theta) &=& \frac{\sum_{ij} w_i w_j (\epsilon_{t, i}^{\mathrm{in}}\epsilon_{ t, j}^{\mathrm{in}} \pm \epsilon_{\times , i}^{\mathrm{in}}\epsilon_{\times , j}^{\mathrm{in}}) \Delta_\theta(ij)}{N_p(\theta)}\ , \nonumber \\
 \hat \xi_\pm^{ss}(\theta) &=& \frac{\sum_{ij} w_i w_j (\gamma_{t,i}\gamma_{t,j} \pm \gamma_{\times , i}\gamma_{\times , j}) \Delta_\theta(ij)}{N_p(\theta)}\ ,\nonumber \\
 \hat \xi_\pm^{sn}(\theta) &=& \frac{\sum_{ij} w_i w_j (\epsilon_{t, i}^{\mathrm{in}}\gamma_{t,j} \pm \epsilon_{\times, i}^{\mathrm{in}}\gamma_{\times , j}) \Delta_\theta(ij)}{N_p(\theta)}
 \end{eqnarray}
 which are the autocorrelation of the intrinsic shape noise, the autocorrelation of the shear signal and their cross correlation. The whole estimator \ref{eq:xi_pm_estimator} is given in terms of these as
 \begin{equation}
  \hat \xi_\pm(\theta) =  \hat \xi_\pm^{nn}(\theta)  +  \hat \xi_\pm^{ss}(\theta)  + 2\cdot \hat \xi_\pm^{sn}(\theta) \ .\nonumber
 \end{equation}
Under the assumption that the shear signal and the shape noise are independent of each other it is obvious that
\begin{equation}
\langle\hat \xi_\pm^{nn}(\theta_1)\hat \xi_\pm^{sn}(\theta_2)\rangle = 0= \langle\hat \xi_\pm^{ss}(\theta_1)\hat \xi_\pm^{sn}(\theta_2)\rangle\ . \nonumber
\end{equation}
If the intrinsic shape of any two galaxies is assumed to be uncorrelated, we can also conclude that
\begin{equation}
\langle\hat \xi_\pm^{nn}\rangle = 0\ \mathrm{for}\ \theta > 0
\end{equation}
and hence
\begin{equation}
\langle\hat \xi_\pm^{nn}(\theta_1)\hat \xi_\pm^{ss}(\theta_2)\rangle = \langle\hat \xi_\pm^{nn}(\theta_1)\rangle\cdot \langle\hat \xi_\pm^{ss}(\theta_2)\rangle = 0 \ \mathrm{for}\ \theta_1 , \theta_2 > 0\ . \nonumber
\end{equation}
 The covariance matrix can thus be split into three different contributions,
\begin{equation}
C_{\pm, \pm}= C_{\pm, \pm}^{nn} + C_{\pm, \pm}^{ss} + C_{\pm, \pm}^{sn}\ , 
\end{equation}
namely
\begin{eqnarray}
\label{eq:cov_contributions}
C_{\pm, \pm}^{nn}(\theta_1, \theta_2) &=& \langle\hat \xi_\pm^{nn}(\theta_1)\hat \xi_\pm^{nn}(\theta_2)\rangle\ , \nonumber \\
C_{\pm, \pm}^{ss}(\theta_1, \theta_2) &=& \langle\hat \xi_\pm^{ss}(\theta_1)\hat \xi_\pm^{ss}(\theta_2)\rangle - \xi_\pm(\theta_1)\xi_\pm(\theta_2)\ , \nonumber \\
C_{\pm, \pm}^{sn}(\theta_1, \theta_2) &=& 4 \cdot \langle\hat \xi_\pm^{sn}(\theta_1)\hat \xi_\pm^{sn}(\theta_2)\rangle\ .
\end{eqnarray}
The $C_{\pm, \pm}^{ss}$ term depends on 4-point functions of the shear field and is called the \emph{cosmic variance} term. In order to evaluate it, further assumptions on the probability distribution function (PDF) of the shear or the convergence field are needed and we will discuss two possible models for the convergence PDF in sections \ref{sec:Gaussian_PDF} and \ref{sec:lognormal_PDF} - the Gaussian and the log-normal model.

The contributions $C_{\pm, \pm}^{nn}$ and $C_{\pm, \pm}^{sn}$ can be computed without additional assumptions. In \citet{joa2008} it is derived that they are given by\footnote{as in \cite{Schneider2002} they employ an ensemble average over the galaxy positions to derive their expressions.}
\begin{eqnarray}
\label{eq:spatial_sn_and_nn}
C_{\pm\pm}^{sn}(\theta_1, \theta_2) &=& \frac{\sigma_\epsilon^2}{\pi A\bar n}\int \mathrm{d}\ell\ \ell\ J_{0/4}(\ell\theta_1) \ J_{0/4}(\ell\theta_2) \ P_\kappa(\ell)\ , \nonumber \\
C_{++}^{nn}(\theta_1, \theta_2) &=& C_{--}^{nn}(\theta_1, \theta_2)\nonumber \\
&=& \frac{\sigma_\epsilon^4}{N_p(\theta_1)}\cdot \delta_{\theta_1,\theta_2}\ , \nonumber \\
C_{+-}^{nn}(\theta_1, \theta_2) &=& 0\ ,
\end{eqnarray}
where $A$ is the survey area, $\bar n$ is the number density of galaxies, $\sigma_\epsilon$ is the dispersion of the intrinsic ellipticity which is defined as
\begin{equation}
\sigma_\epsilon^2 := \langle\boldsymbol\epsilon^{\mathrm{in}} {\boldsymbol{\epsilon}^{\mathrm{in}}}^*\rangle\ ,
\end{equation}
and $P_\kappa$ is again the convergence power spectrum.

\subsubsection{Gaussian Approximation}
\label{sec:Gaussian_PDF}
In the paper series by \citet{Schneider2002}, \citet{Kilbinger2} and \citet{joa2008} the covariance matrix is studied in the Gaussian approximation, i.e. assuming that the convergence field has a Gaussian PDF such that its 4-point correlation functions can be expressed in terms of its 2-point correlation functions. 

For the case where the survey geometry is much larger than the angular scales considered in the correlation functions, \citet{joa2008} derive the following expressions for the cosmic variance term:
\begin{equation}
\label{eq:joachimi_covariance}
C_{\pm\pm}^{ss}(\theta_1, \theta_2) = \frac{1}{\pi A} \int \mathrm{d}\ell\ \ell\ J_{0/4}(\ell\theta_1) \ J_{0/4}(\ell\theta_2) \ P_\kappa^2(\ell)\ .
\end{equation}
However, due to the finite geometry of any given survey equation \ref{eq:joachimi_covariance} generally overestimates the covariance of Gaussian field as was demonstrated in \citet{Sato2011}. This \emph{finite area effect} according to \citeauthor{Sato2011} is not important for surveys larger than $1000\deg^2$. For smaller surveys a method developed in \citet{Kilbinger2} which doesn't employ an ensemble average over galaxy positions should be used to evaluate the Gaussian covariance. This method was for example used in the analysis of CHFTLenS data in \citet{Kilbinger1}. The finite area effect is also important for internal covariance estimation and will be further discussed in section \ref{sec:subsample_correlation}.

\subsubsection{Shifted Log-Normal Approximation}
\label{sec:lognormal_PDF}

As e.g. reported by \citet{Taruya2002}, \citet{Vale2003} or by \citet{Hilbert2011} the Gaussian model fails to describe the true PDF of the convergence and especially on small separations poorly represents the true covariance of the cosmic shear 2-point functions.

\citet{Hilbert2011} propose a different model for the convergence PDF, namely that of a \emph{zero-mean shifted log-normal distribution}. In this approach the convergence at a given point on the sky is assumed to be of the form
\begin{equation}
\label{eq:kappa_n}
\kappa(\boldsymbol \theta) = \exp[n(\boldsymbol \theta)] - \kappa_0
\end{equation}
where $n(\boldsymbol \theta)$ is a Gaussian random field (not necessarily with a vanishing mean) and the \emph{minimal convergence parameter} $\kappa_0$ is chosen such that $\langle\kappa\rangle = 0$. \citet{Hilbert2011} show that from the corresponding PDF a model for the shear-shear contribution to the covariance matrix can be derived. Considering only the most important terms they also provide a simplified log-normal covariance, which reads
\begin{eqnarray}
\label{eq:log-norm_covariance}
C_{\pm\pm}^{ss}(\theta_1, \theta_2) &=& \frac{1}{\pi A} \int \mathrm{d}\ell\ \ell\ J_{0/4}(\ell\theta_1) \ J_{0/4}(\ell\theta_2) \ P_\kappa^2(\ell) \nonumber \\
&& + \frac{8\pi}{\kappa_0^2 A}\ \xi_\pm(\theta_1) \xi_\pm(\theta_2) \int_0^{\theta_\mathrm{A}} \mathrm{d}\theta \ \theta \ \xi_+(\theta) \ ,
\end{eqnarray}
where $\theta_A$ represents the 'radius' of the survey, given by
\begin{equation}
\theta_\mathrm{A} = \sqrt{\frac{A}{\pi}}\ .
\end{equation}
Comparing equation \ref{eq:log-norm_covariance} to equation \ref{eq:joachimi_covariance} on can see that the simplified log-normal approximation to $C_{\pm\pm}^{ss}$ consists of only one correction term to the Gaussian model. In our paper, we will simulate log-normally distributed convergence fields and use equation \ref{eq:log-norm_covariance} to compute the cosmic variance part of our model covariance.

\subsubsection{Finite bin width}
\label{sec:finite_bins}

The expressions presented above for the covariance of $\hat \xi_\pm$ are derived under the assumption of small angular bins \citep{Schneider2002}. However, in section \ref{sec:des_sv} we need correct covariance expressions also for data vectors where the relative bin width is $\sim 0.3$, i.e. where the assumption of small bins does not hold. This is in fact the more realistic case, since broad bins are commonly used to reduce the number of data points (see e.g. \citet{Kilbinger1}, \citet{Becker2015}).

Hence, in section \ref{sec:des_sv} we proceed as follows: We first compute the log-normal model for the covariance, eqn. \ref{eq:log-norm_covariance}, for a set of very small angular bins $\tilde{\theta}_i, i = 1, \dots , \tilde{N}$. Then we apply a linear transformation that takes the large data vector of the small angular bins to a smaller data vector by putting together $p$ neighbouring bins of the old data vector,
\begin{eqnarray}
\label{eq:finite_bin_covariance}
\theta_j &=& \sum_{i=p\cdot (j-1) + 1}^{p\cdot j} \tilde{\theta}_i N_p(\tilde{\theta}_i) / \sum_{i=p\cdot (j-1) + 1}^{p\cdot j} N_p(\tilde{\theta}_i) \nonumber \\
\hat\xi(\theta_j) &=& \sum_{i=p\cdot (j-1) + 1}^{p\cdot j} \hat\xi(\tilde{\theta}_i) N_p(\tilde{\theta}_i) / \sum_{i=p\cdot (j-1) + 1}^{p\cdot j} N_p(\tilde{\theta}_i)\ , \nonumber \\
\end{eqnarray}
where $N_p(\tilde{\theta}_i)$ is the number of pairs in the $i$th bin of the finer data vector.

The same linear transformation is then applied to the covariance matrix of the large data vector to get the covariance matrix of the compressed data vector. We find that for $\hat \xi_- $ this decreases the mixed- and cosmic variance part of the covariance by $\gtrsim 30\%$, while for $\hat \xi_+ $ it makes almost no difference.  The reason is that adjacent bins in $\xi_+$ are much more correlated than adjacent bins in $\xi_-$. Hence, if two bins in $\xi_+$ are joined, the variance of the joined bin is almost identical to that of the individual bins and eqn. \ref{eq:log-norm_covariance} can still be applied\footnote{a similar reasoning can be applied for the off-diagonal terms of the covariance}.

\section{Log-normal Simulations}
\label{sec:log-normal_simulations}

\citet{Simon2004} describe a quick method to simulate cosmic shear surveys based on a Gaussian convergence field for any given convergence-power-spectrum. On a quadratic grid in 2D-Fourier space they generate at each point $\boldsymbol \ell$ of the grid a value of the convergence
\begin{equation}
\hat \kappa(\boldsymbol \ell) = \kappa_1(\boldsymbol \ell) + i \kappa_2(\boldsymbol \ell) \nonumber
\end{equation}
where the components $\kappa_i(\boldsymbol \ell)$ are drawn from a Gaussian distribution with zero mean and variance
\begin{equation}
\sigma_\ell^2 = \frac{1}{2V} P_\kappa(\ell). \nonumber
\end{equation}
Here $P_\kappa$ is the desired convergence power-spectrum and $V$ is the volume of the grid in angular space which is given in terms of the grid spacing $\Delta \ell$ as
\begin{equation}
\label{eq:angular_volume}
V = \left( \frac{2\pi}{\Delta \ell} \right)^2\ .
\end{equation}
In order to achieve a convergence field that is real valued in angular space one has to impose the condition 
\begin{equation}
\hat \kappa(\boldsymbol \ell) = \hat \kappa^*(-\boldsymbol \ell)  \nonumber
\end{equation}
and in Fourier space the shear field is related to the convergence field by the equation\footnote{see eqn. 2.1.11 of \citet{Kaiser1993} or eqn. 25 of \citet{Simon2004}}
\begin{equation}
\label{eq:kappa_to_gamma}
\hat \gamma(\boldsymbol \ell) = \frac{\ell_1^2-\ell_2^2+2i\ell_1\ell_2}{\ell^2} \hat \kappa(\boldsymbol \ell)\ .
\end{equation}
A Fourier transform then gives the shear field in angular space.

The main idea in generating a log-normal random field is to generate a Gaussian field $n(\boldsymbol\theta)$ with the method of \citet{Simon2004} and transform it into $\kappa(\boldsymbol\theta)$ via equation \ref{eq:kappa_n}. According to \citet{Martin2012, Takahashi2014} the power spectrum of $n(\boldsymbol\theta)$, $P_n$, can be computed from $P_\kappa$ as follows: 

related to the 2-pt. function of $n(\boldsymbol\theta)$ via (see e.g. equation B.8 of \citealt{Hilbert2011}):

First, the 2-pt. function of $\kappa(\boldsymbol\theta)$ is given in terms of the power spectrum $P_\kappa$ by
\begin{equation}
\xi_\kappa(\theta) = \frac{1}{2\pi}\int_0^\infty\ d\ell\ \ell\ P_\kappa(\ell)\ J_0(\ell\theta)\ . \nonumber \end{equation}
Next, the 2-pt. function $\xi_\kappa$ is related to the 2-pt. function of $n(\boldsymbol\theta)$ via (see e.g. equation B.8 of \citealt{Hilbert2011})
\begin{equation}
\xi_n(\theta) = \ln \left(\xi_\kappa(\theta) / \kappa_0^2 + 1  \right)\ , \nonumber
\end{equation}
where $\kappa_0$ is the minimal convergence parameter from eqn. \ref{eq:kappa_n}. Finally, the power spectrum of the Gaussian field $n(\boldsymbol\theta)$ by
\begin{equation}
P_n(\ell) = 2\pi\int_0^\infty\ d\theta\ \theta\ \xi_n(\theta)\ J_0(\ell\theta)\ .
\end{equation}

The field $n(\boldsymbol\theta)$ can now be generated as described by \citet{Simon2004}. However, this way $n(\boldsymbol\theta)$ will have a mean value of zero. In order to ensure that $\langle \kappa \rangle = 0$ the mean value
\begin{equation}
\mu = \kappa_0 - \frac{\sigma^2}{2}
\end{equation}
has to be added, where $\sigma^2$ is the variance of the Gaussian field. The convergence field $\kappa(\boldsymbol\theta)$ now has to be transformed into Fourier space. Using equation \ref{eq:kappa_to_gamma} one can then compute the Fourier modes of the shear field and another Fourier transform gives the desired shear field in angular space.

\subsection{Setup and Validation of the Simulations}
\label{sec:setup}

\textcolor{blue}{
The harmonic space grid we are using has a total number of $(2^{16})^2$ grid points and a grid spacing of $\Delta \ell = 2$. Hence, in each axis it ranges from $-\ell_{\max} = -2^{16}$ to $\ell_{\max} = +2^{16}$. All modes $\boldsymbol{\gamma}(\boldsymbol{\ell})$ with $|\boldsymbol{\ell}| > \ell_{\max}$ (i.e. the corners of the grid) are set to zero. The mode $\boldsymbol{\gamma}_0$ is also set to $0$ and all other modes are generated as explained above. Especially, we have to fix a cosmology and assume a certain redshift distribution of sources, $p(z)$, to compute the convergence power spectrum $P_\kappa$. }

\textcolor{blue}{
Following eqn. \ref{eq:angular_volume} the grid in angular space has a volume of $V = 2\pi / 2 \approx 10^4 \deg^2$. Out of the center of that volume we will cut out a sub-grid of size $A$. Onto that sub-grid we are uniformly placing galaxies with a certain number density $n_{\mathrm{gal}}$. The shear of each individual galaxy is then determined by quadratic interpolation of the grid onto the galaxy position. Finally, a Gaussian intrinsic shape noise with an ellipticity dispersion $\sigma_{\epsilon}$ is added to get the total shape of the galaxy. Note that we simply added the shear signal and intrinsic ellipticity, hereby ignoring the effects of \emph{reduced shear}. }

\textcolor{blue}{
In this work we always keep the cosmology fixed to that of \citet{Hilbert2011}, i.e. a flat $\Lambda$CDM universe with $(\Omega_m, \Omega_b, \sigma_8 , h_{100}, n_s) = (0.25, 0.045, 0.9, 0.73, 1.0)$. To compute the convergence power spectrum we employ \emph{halofit} \citep{Smith2003} using the open source code \textsc{nicaea}\footnote{by Kilbinger et al., www.cosmostat.org/software/nicaea/}. The source distribution is taken to have the form
\begin{equation}
\label{eq:p_of_z}
p(z) = \frac{3z^2}{2z_0^3}\ e^{-\left(\frac{z}{z_0}\right)^{3/2}}\ ,\ \mathrm{where}\ z_0 = \frac{z_{\mathrm{median}}}{1.412}\ .
\end{equation}
This is the same form that was also assumed by \citet{Hilbert2011}. The ellipticity dispersion is always set to $0.3$ per component, i.e. $\sigma_{\epsilon} = \sqrt(2) \cdot 0.3$. All other quantities, i.e. area $A$, source density $n_{\mathrm{gal}}$ and median redshift $z_{\mathrm{median}}$, will be varied throughout section \ref{sec:jackknife_results}. The different setups are summarized in table \ref{tab:setup}.}

\textcolor{blue}{
The redshift distribution of setup I is exactly that of \citet{Hilbert2011} and imitates a rather deep survey comparable e.g. to euclid. In this setup, we measure the 2-pt. correlation functions in $35$ logarithmic bins from $\theta_{\min} = 1'$ to $\theta_{\max} = 150'$. The area $A$ was taken to be a square of $70 \deg \times 70 \deg$. The minimal convergence parameter $\kappa_0$ was chosen to be $0.032$ as suggested by \citeauthor{Hilbert2011} for this redshift distribution.}

\textcolor{blue}{
The area, galaxy density and redshift distribution of setup IIa are chosen to be similar to that of DES science verification data (DES-SV) which was used in \citet{Becker2015}. In this setup, we measure the 2-pt. correlation functions in $15$ logarithmic bins from $\theta_{\min} = 2'$ to $\theta_{\max} = 300'$, which is also exactly the data vector used by \citet{Becker2015}. In this setup we also reproduce the irregular shape of DES-SV, i.e. we use an SV-shaped healpix mask to cut out the sub-volume $A$.}

\textcolor{blue}{
The setups IIb and IIc are aimed at a forecast for the final 5-year DES data. In IIb we are assuming the same source density as in DES-SV and in IIc a slightly higher one. Note that in principle, when adjusting the source density, we should also adjust the source median redshift of the sources. But we will ignore this point, since our redshift distribution is anyway only a rough match to that of DES. Thus, for all setups IIa, IIb and IIc we take a median redshift of $0.7$. Furthermore, for all these setups we use an empirical formula $\kappa_0(z)$ found by \citet{Hilbert2011} to fix the minimal convergence parameter. Inserting the mean redshift of $z_{\mathrm{mean}} \approx 0.745$ gives a value of $\kappa_0 = 0.019$. The area in setups IIb and IIc are simply taken to be square shaped.}

\begin{table}
\begin{center}
\begin{tabular}{ c || c | c | c  | c } 
setup & $A\ [\deg]$ & $n_{\mathrm{gal}}$ & $z_{\mathrm{median}}$ & $\kappa_0$\\ 
\hline
I & $4900$  & $20$ & $1.0$ & $0.032$ \\
IIa  & $\sim 150$ & $6$ & $0.7$ & $0.019$ \\
IIb  & $5000$ & $6$ & $0.7$ & $0.019$ \\
IIc  & $5000$ & $10$ & $0.7$ & $0.019$ \\
\end{tabular}
\end{center}
\caption{The different configurations of mock catalogs used in this paper.}
 \label{tab:setup}
\end{table}

\begin{figure}
\begin{centering}
\includegraphics[width=0.49\textwidth]{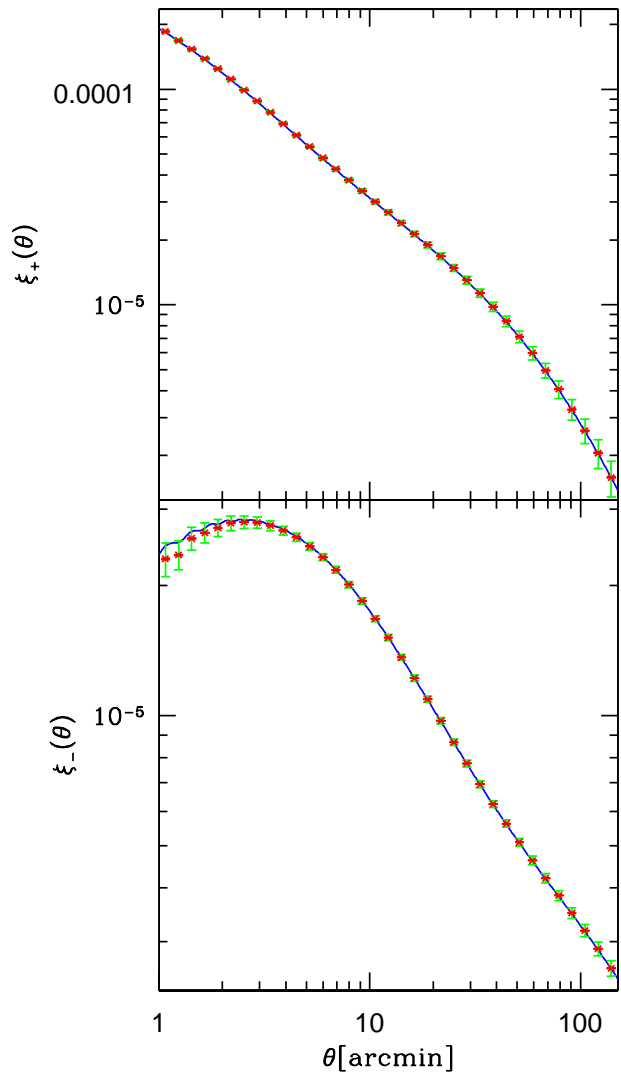}
\end{centering}
 \caption{Comparison of the mean correlation functions from $1000$ simulations (red dots) and the input model (blue line). The the red error bars show the standard deviation of the mean and the green errorbars show the standard deviation of the single correlation function measurements. We used the redshift distribution of \citet{Hilbert2011} to compute the input power spectrum and we also used their value of $\kappa_0$ to generate the log-normal convergence. Note that in section \ref{sec:des_sv} we will use a different configuration.}
\label{fig:xi_1000}
\end{figure}

\textcolor{blue}{
To validate our simulations, we generate 1000 independent realizations of setup I. In order to speed up the computations we decrease the number of galaxies with respect to our jackknife analysis by a factor of $5$, i.e. to $n_\mathrm{gal} = 4/{\mathrm{arcmin}}^2$, while at the same time decreasing the ellipticity dispersion by a factor of $\sqrt{5}$. This way the covariance expressions in equation \ref{eq:log-norm_covariance} stay unaffected. }

\begin{figure*}
\begin{centering}
\includegraphics[width=0.49\textwidth]{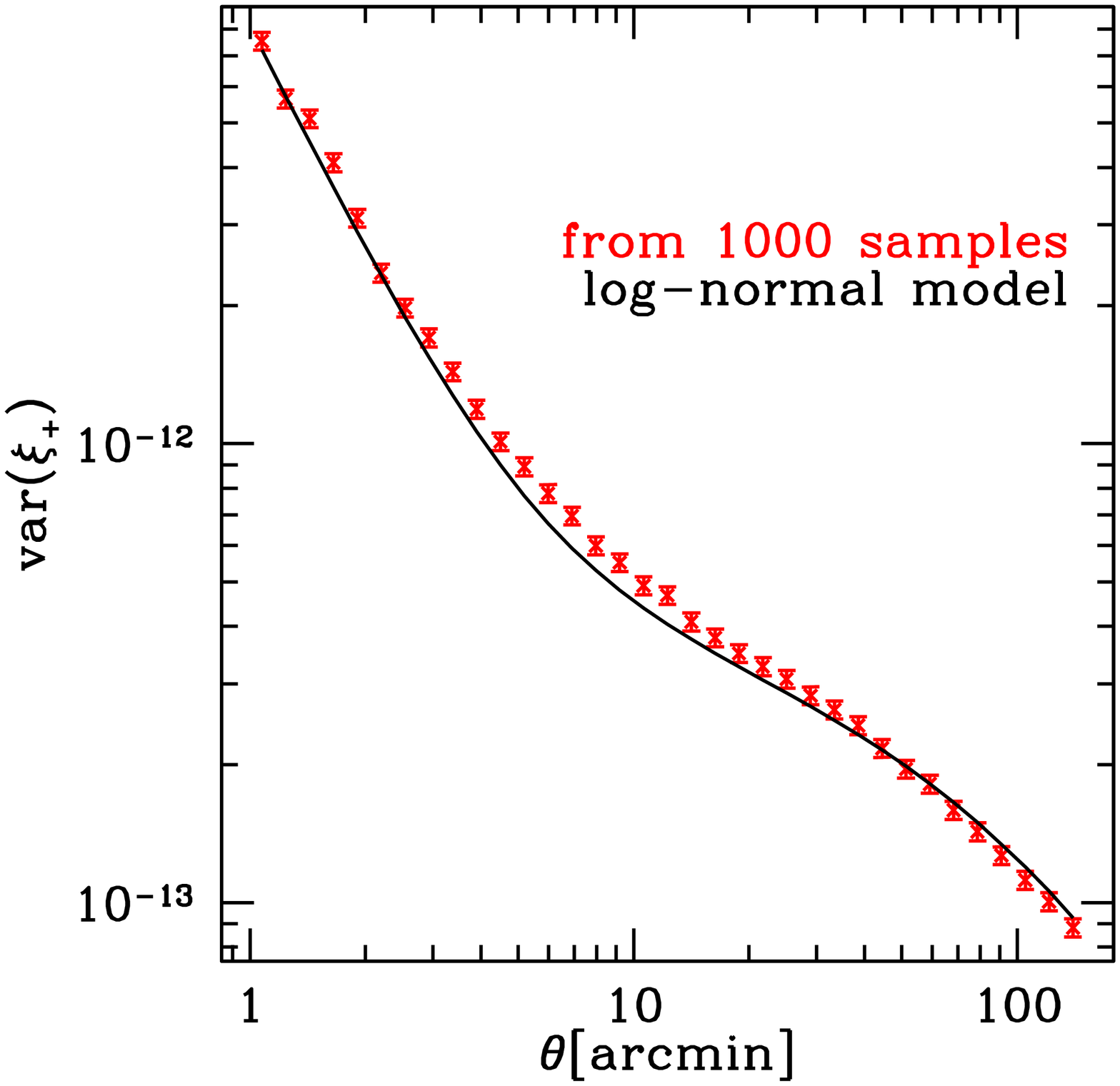}\includegraphics[width=0.49\textwidth]{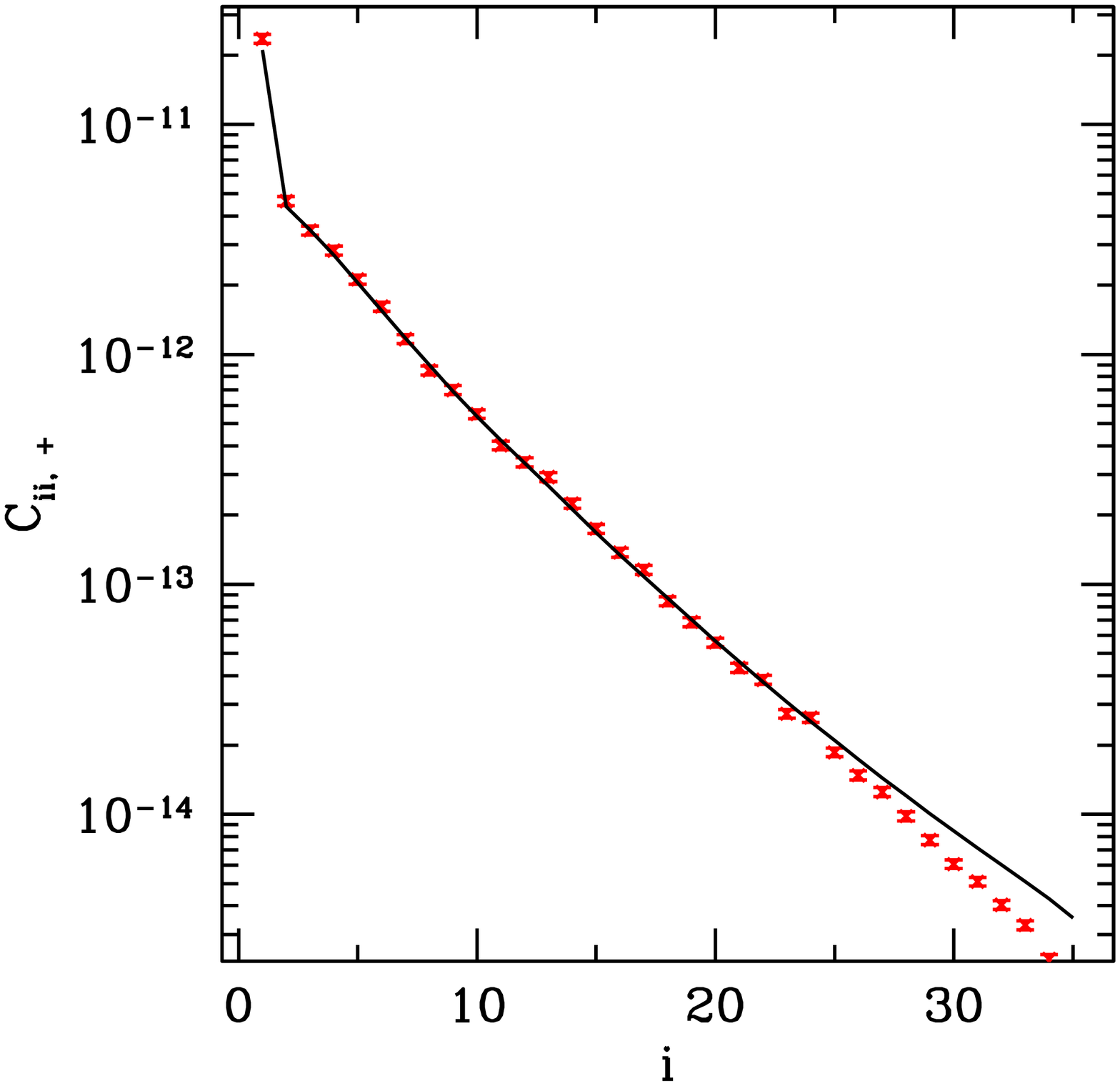}

\includegraphics[width=0.49\textwidth]{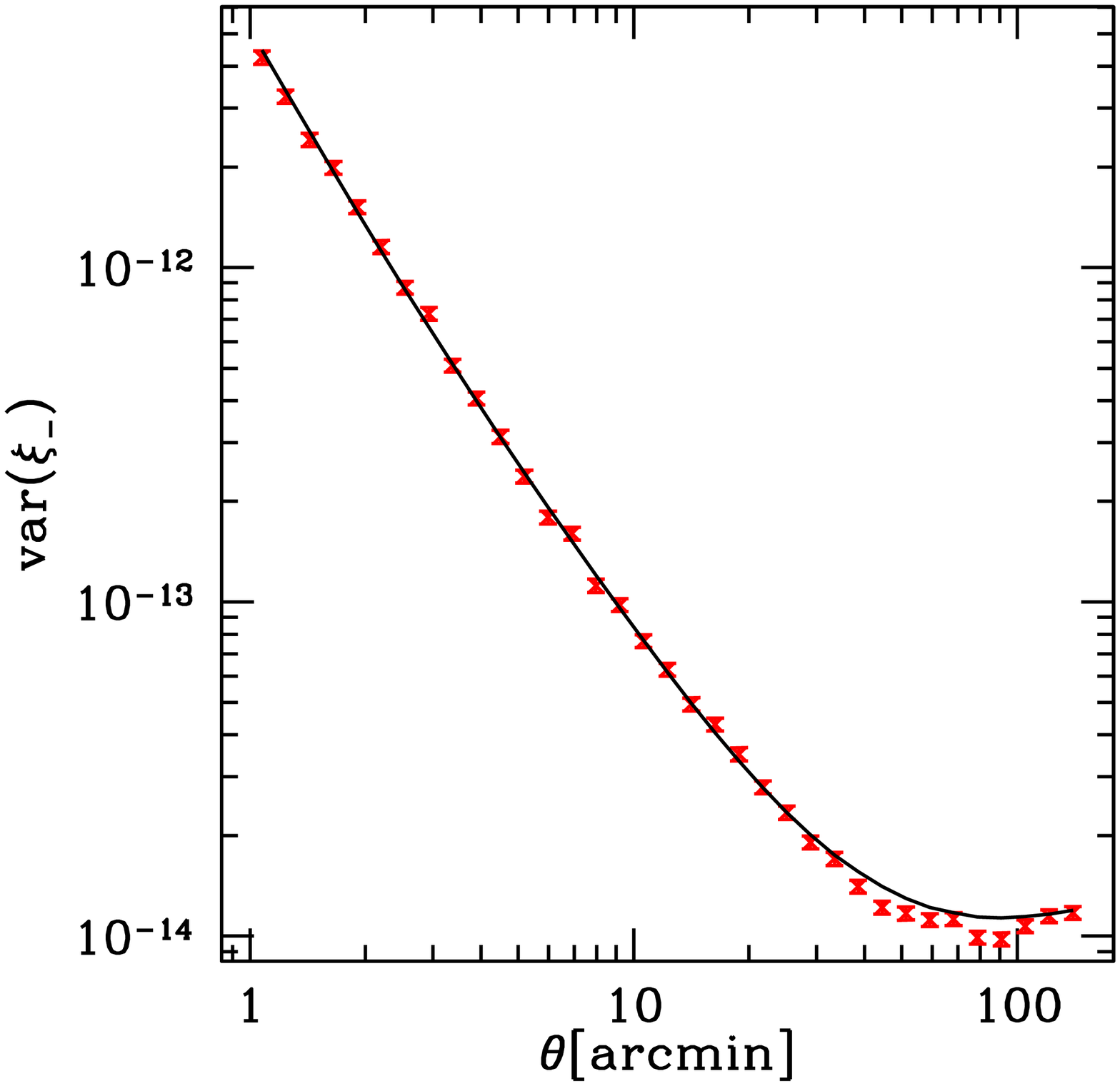}\includegraphics[width=0.49\textwidth]{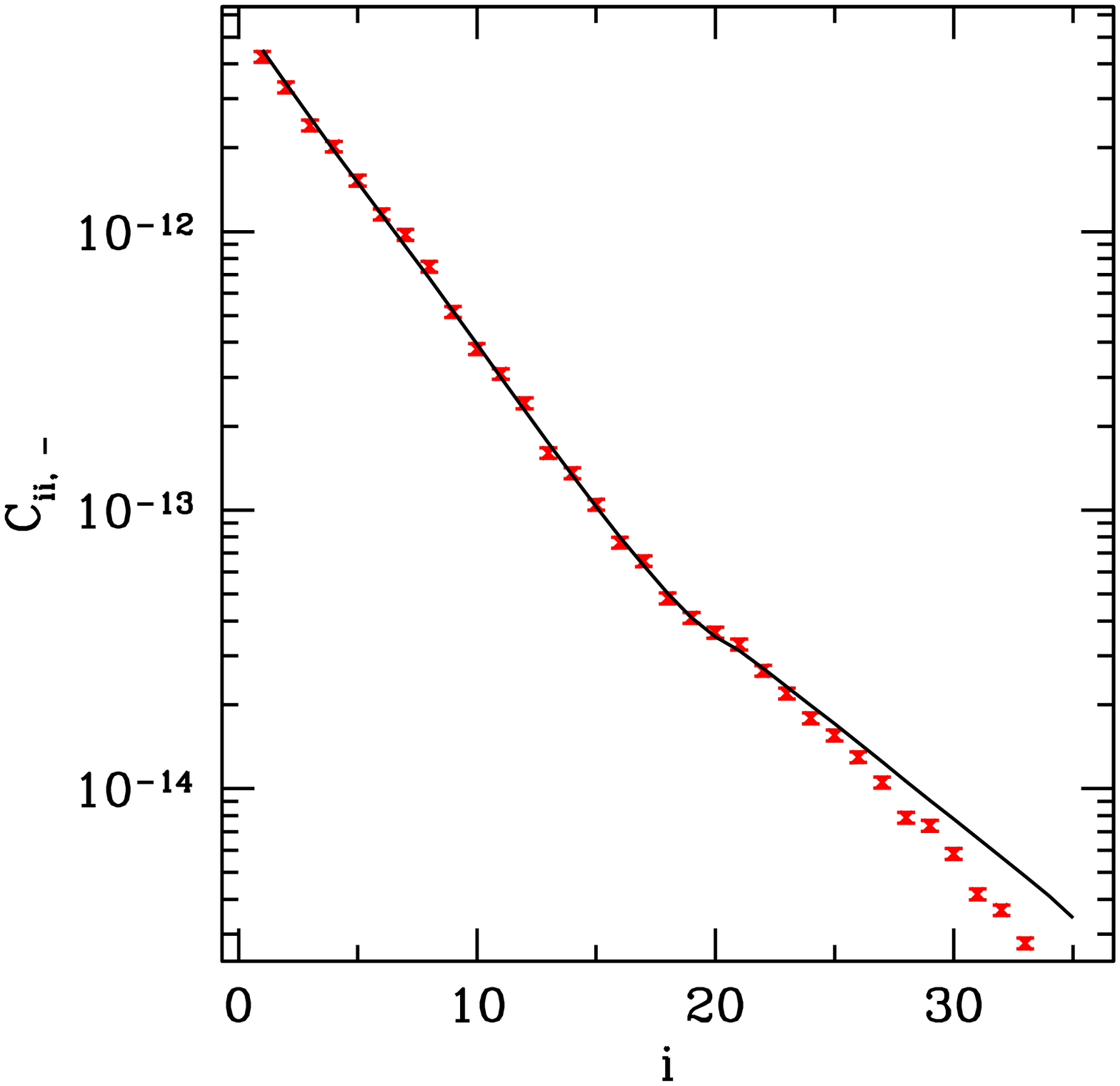}
\end{centering}
 \caption{Left: sample variance from $1000$ independent simulations compared to the log-normal input model. The errorbars are assuming a Wishart distribution, note however that
 the different sample variance values are correlated.
 Right: in the diagonal basis of the model covariance matrix the sample variance values should independently follow a $\chi^2$-distribution. The model and the simulations are consistent for the $\approx 20$ largest eigenvalues of the model covariance matrix.}
\label{fig:var_1000}
\end{figure*}

\textcolor{blue}{
In figure \ref{fig:xi_1000} we show the mean measured correlation functions in the mock surveys. The correlation function measurement was carried out using the \textsc{TreeCorr} tree code\footnote{by Jarvis et al., github.com/rmjarvis/TreeCorr}. The measured correlation functions and those derived from the input model agree well on most scales. Only at small angular scales the measured value of $\xi_-$ differs significantly from the input model. The reason is the artificial cut-off at high $\ell$-values in our Fourier grid which both in the model and the simulation introduces artefacts -  as can be seen from the oscillatory behaviour of $\xi_-$. To keep our analyses in section \ref{sec:jackknife_results} free from these artefacts we will only consider those bins in $\xi_-$ that have $\theta \gtrsim 4.5'$. For $\xi_+$ we continue to use a range of $1' < \theta < 150'$. Also, for the setups $IIa$ to $IIc$ (not shown here) the discrepancy in $\xi_-$ turns out to be less significant. Hence for these setups we stay with $\theta_{\min} = 2'$.}

Figure \ref{fig:var_1000} compares the sample covariance of the $1000$ simulations to the predictions from equation \ref{eq:log-norm_covariance}. The relative deviation between measured variance and the log-normal model is $\leq 20\%$ for $\xi_+$  and $\leq 15\%$ for $\xi_-$. For both correlation functions these deviations seem to be significant given the uncertainties of the sample covariance estimate. However, the sample variance values at different angular scales are highly correlated, which makes a '$\chi$-by-eye' judgement of the fit impossible. When transforming the covariance matrices into the eigenbasis of the model covariance (right-hand panel of figure \ref{fig:var_1000}), the variance values become uncorrelated and the agreement of the covariance matrices becomes more evident. The eigenvalues at which the log-normal covariance significantly differs from the sample covariance of our simulations are $3$ orders of magnitude smaller than the biggest eigenvalues for $\xi_+$ and more than $2$ orders of magnitude smaller than the biggest eigenvalues for $\xi_-$ (c.f. right-hand panel of figure \ref{fig:var_1000}). Finally, our analyses in section \ref{sec:jackknife_results} remain unchanged when the log-normal covariance is exchanged by the sample covariance of the $1000$ independent realizations, which validates the simulations for our purposes (cf. appendix \ref{sec:appendix_off_diagonal}, figure \ref{fig:constraints_empirical_covariance}).

\section[]{Internal Covariance Estimation for two-point correlation functions}
\label{sec:jackknife_schemes}

Suppose the correlation functions $\xi_\pm$ have been measured in finite bins around a set of angular distances $\theta_i,\ i = 1, \dots , d$. Let $\hat{\boldsymbol \xi}$ be either one of the data vectors $[\xi_\pm(\theta_1), \dots , \xi_\pm(\theta_d) ]$ or the joint data vector of both correlation functions.

If $\boldsymbol \xi[\boldsymbol \pi]$ is a model for the measurement $\hat{\boldsymbol \xi}$ which depends on a set of parameters $\boldsymbol \pi$, then a common statistic for constraining the possible values of $\boldsymbol \pi$ is the $\chi^2$ statistic \citep{Kilbinger2}, i.e.
\begin{equation}
\label{eq:figure_of_merit}
\chi^2[\boldsymbol \pi] = (\hat{\boldsymbol \xi} - \boldsymbol \xi[\boldsymbol \pi])^T C^{-1}\ (\hat{\boldsymbol \xi} - \boldsymbol \xi[\boldsymbol \pi])\ ,
\end{equation}
where $C$ is the covariance matrix of $\hat{\boldsymbol \xi}$. One way to get the covariance matrix is to model it theoretically. As we have seen in section \ref{sec:covariance_models} the modelling of the covariance depends crucially on the PDF of the convergence field \citep{Schneider2002, Hilbert2011, Sato2011} and neither the Gaussian nor the log-normal approximation match a realistic convergence PDF. Also, the model covariance matrix will depend on cosmological parameters itself which, at least for small surveys, has to be taken into account when deriving parameter constraints \citep{Eifler2009}.

A way to get around modelling the covariance matrix directly is to use the sample covariance of measurements of the correlation functions in a set of independent N-body simulations (cf. \citealt{Takahashi2009, Sato2009, Harnois2015} or for an application to data \citealt{Kilbinger1}) which however still depends on the model parameters, i.e. on the assumption of a particular cosmological model. Another alternative to modelling the covariance matrix is to estimate it from the data itself. In the following we will introduce three different internal covariance estimation methods - the \emph{sub-sample covariance}, the \emph{delete-one-jackknife} and the \emph{bootstrap} (cf. \citealt{Norberg2009, Loh2008}).

\subsection{Subsample Covariance}

Let us split the area $A$ of our cosmic shear survey into $N$ equally shaped and sized subregions of the area $A_{\mathrm{S}} = A/N$. In each subregion $\alpha = 1, \dots , N$, a measurement of the data vector $\hat{\boldsymbol \xi}^\alpha$ can be carried out. Assuming that each sub-region has approximately the same number of galaxies and that the correlation functions are measured on scales much smaller than $\sqrt{A_{\mathrm{S}}}$ the measurement of $\hat{\boldsymbol \xi}$ in the whole survey is given by
\begin{equation}
\hat{\boldsymbol \xi} \approx \bar{\boldsymbol \xi} := \frac{1}{N} \sum_{\alpha = 1}^N  \hat{\boldsymbol \xi}^\alpha\ ,
\end{equation}
i.e. it is the mean values of the measurements in the sub-regions. If the measurements $\hat{\boldsymbol \xi}^\alpha$ are independent, then the $ij$-th element of their covariance matrix can be estimated by
\begin{equation}
\langle \Delta\hat{\boldsymbol \xi}_i^\alpha\Delta\hat{\boldsymbol \xi}_j^\alpha \rangle \approx \frac{1}{N-1} \sum_{\beta = 1}^N (\hat{\boldsymbol \xi}^\beta- \boldsymbol{\bar\xi})_i\ (\hat{\boldsymbol \xi}^\beta- \boldsymbol{\bar\xi})_j\ ,
\end{equation}
where $\Delta\hat{\boldsymbol \xi}^\alpha$ is the difference between $\hat{\boldsymbol \xi}^\alpha$ and its expectation value
\begin{equation}
\boldsymbol \xi = \langle\hat{\boldsymbol \xi}^\alpha \rangle = \langle\hat{\boldsymbol \xi} \rangle\ .
\end{equation}
Accordingly, if the assumption of independent sub-regions were true, the covariance of the total measurement $\hat{\boldsymbol \xi}$ could be estimated by
\begin{equation}
\label{eq:subsample_estimator}
\hat C_{\mathrm{SC}} = \frac{1}{N(N-1)} \sum_{\alpha = 1}^N (\boldsymbol \xi^\alpha- \boldsymbol{\bar\xi})^T\ (\boldsymbol \xi^\alpha- \boldsymbol{\bar\xi})\ .
\end{equation}
We will call the estimator in equation \ref{eq:subsample_estimator} the \emph{sub-sample covariance} \citep{Norberg2009}. The main systematic effects of internal covariance estimation can be most easily understood in terms of this estimator. Hence, before introducing the jackknife and bootstrap estimator, we will explain these systematics in the following two sections.

\subsection{Correlation of sub-samples}
\label{sec:subsample_correlation}

The sub-sample covariance estimator relies on the assumption that the data is split into independent sub-samples, i.e. that there is no correlation of the measurements of the correlation functions in the different sub-regions,
\begin{equation}
\langle \Delta\hat{\boldsymbol \xi}_i^\alpha\Delta\hat{\boldsymbol \xi}_j^\beta \rangle \overset{!}{=} 0\ , \ \mathrm{for}\ \alpha \neq \beta\ .
\end{equation}
This can be seen from the fact that eqn. \ref{eq:subsample_estimator} simply rescales the sub-field-to-sub-field covariance by a factor of $1/N$ to get the covariance of the whole survey. If the sub-samples are correlated, this will underestimate the true covariance matrix \citep{Nordmann2007}. 

Another way to think about this is as follows: the sub-sample covariance estimator assumes that the covariance matrix of $\hat{\boldsymbol \xi}$ is inversely proportional to the survey area $A$. Hence it estimates the covariance of sub-regions of the size $A_{\mathrm{S}}$ within the data and then rescales it to the total area,
\begin{equation}
\label{eq:subsample_rescaling}
C = \frac{A_{\mathrm{S}}}{A}\cdot C_{\mathrm{S}} = \frac{1}{N}\cdot C_{\mathrm{S}}\ ,
\end{equation}
where $N$ is again the number of sub-regions. But already from the log-normal model for the covariance it can be seen, that this rescaling is not valid. The log-normal correction term to the Gaussian covariance matrix is given by
\begin{equation}
C_{\pm\pm}^{ss, \mathrm{log}}(\theta_1, \theta_2) = \frac{8\pi}{\kappa_0^2 A}\ \xi_\pm(\theta_1) \xi_\pm(\theta_2) \int_0^{\theta_A} \mathrm{d}\theta \ \theta \ \xi_+(\theta) \ . \nonumber \\
\end{equation}
This term may be proportional to $1/A$, but the upper integral boundary also depends on the survey diameter $\theta_A$. As $A$ increases, the covariance therefore decreases slower than $1/A$. Hence, assuming $1/A$ scaling when extrapolating from the covariance of the smaller sub-fields to the covariance of the full area underestimates the full covariance. Also, note that even the Gaussian covariance term in eqn. \ref{eq:log-norm_covariance} is only an approximation for large survey sizes $A$. It also suffers from a finite area effect as can be seen from its derivation in \citet{Schneider2002} or its form given in \citet{Hilbert2011}.

The fact that sub-samples should be as uncorrelated as possible is also the reason why the re-sampling of the data should be done into spatially connected patches. If instead the data would be randomly divided into sub-samples then the shear correlations in the sub-samples would be almost identical. Hence, only the shape-noise contributions to the covariance would be measured by such an estimator.

\begin{figure}
\centering{
\includegraphics[width=0.35\textwidth]{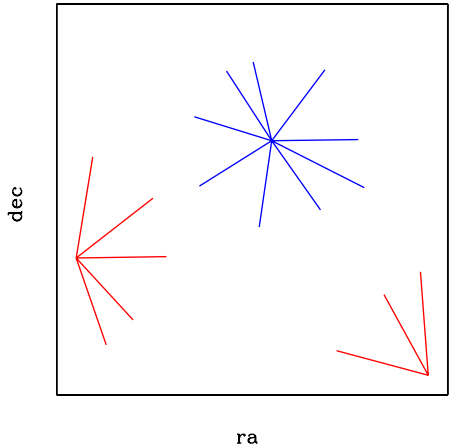}
}
 \caption{Galaxies at the edge of a sub-region (in red) contribute less pairs to the measurement of the correlation functions (i.e. to equation \ref{eq:xi_pm_estimator} applied to the sub-sample) than galaxies in the center of the sub-region (in blue). Consequently, the area of the sub-patch is not uniformly probed by the galaxy pairs. This increases the cosmic variance between sub-regions and biases the covariance estimates high. Hence, it has an opposite effect to the correlation of sub-samples, which biases the covariance estimates low. As seen from the left-hand panel of figure \ref{fig:cov_components}, at large angular scales this can even lead to an overestimation of the cosmic variance of $\hat\xi_-$ (in the galaxy-scheme, c.f. also figure \ref{fig:jack_schemes}).}
\label{fig:effective_area}
\end{figure}

\begin{figure}
\begin{centering}
\includegraphics[width=0.24\textwidth]{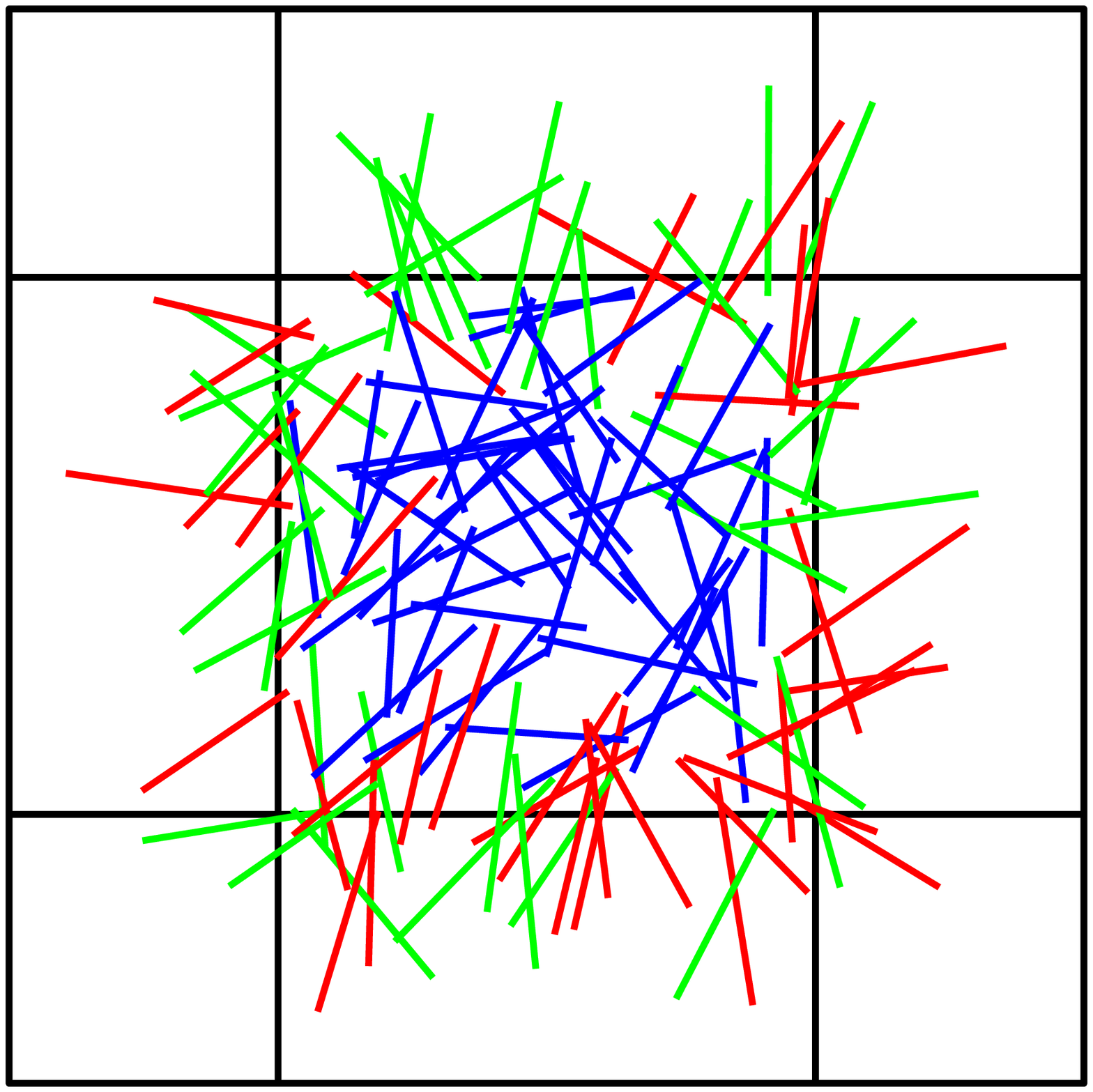}

\includegraphics[width=0.24\textwidth]{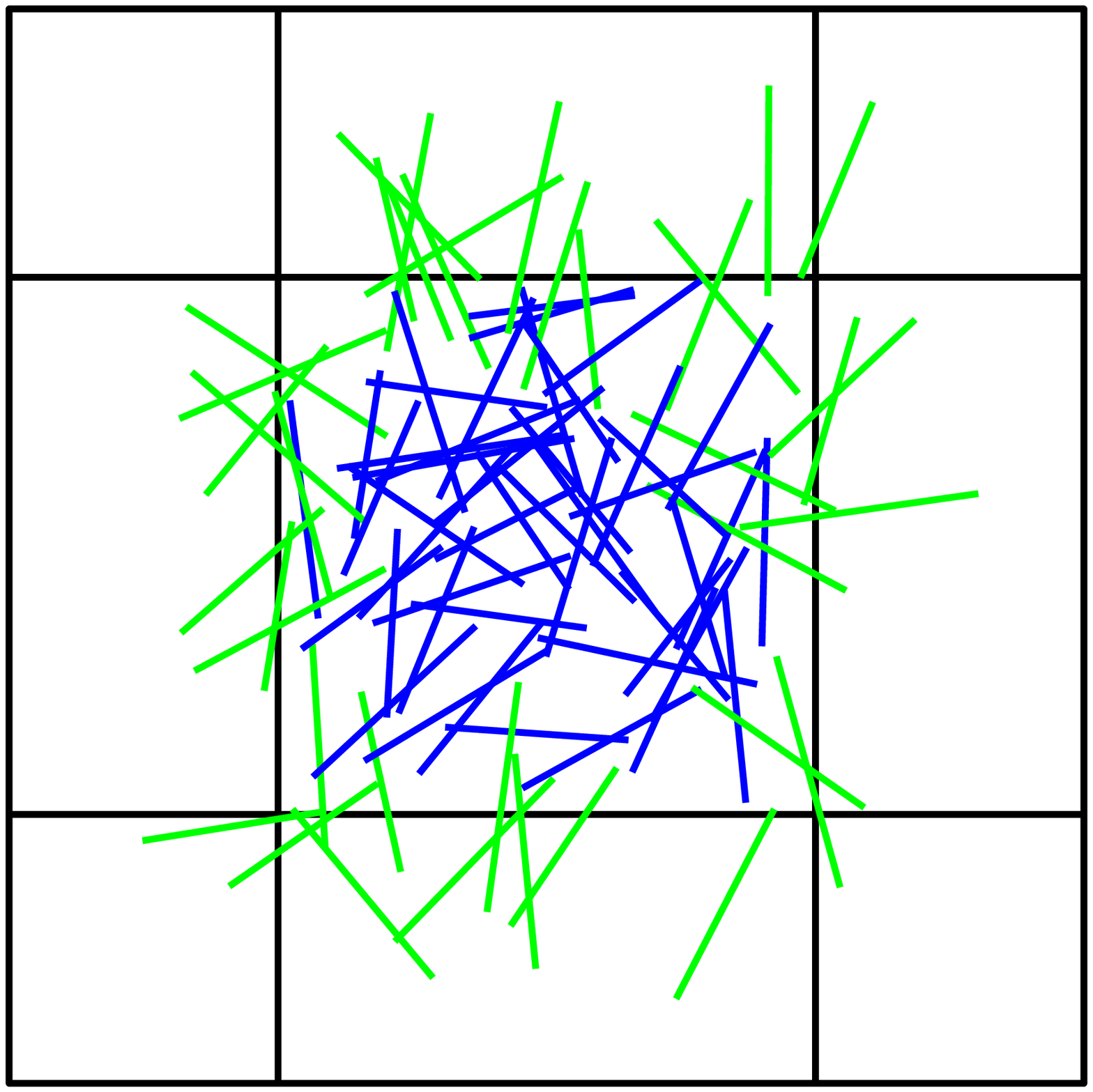}\includegraphics[width=0.24\textwidth]{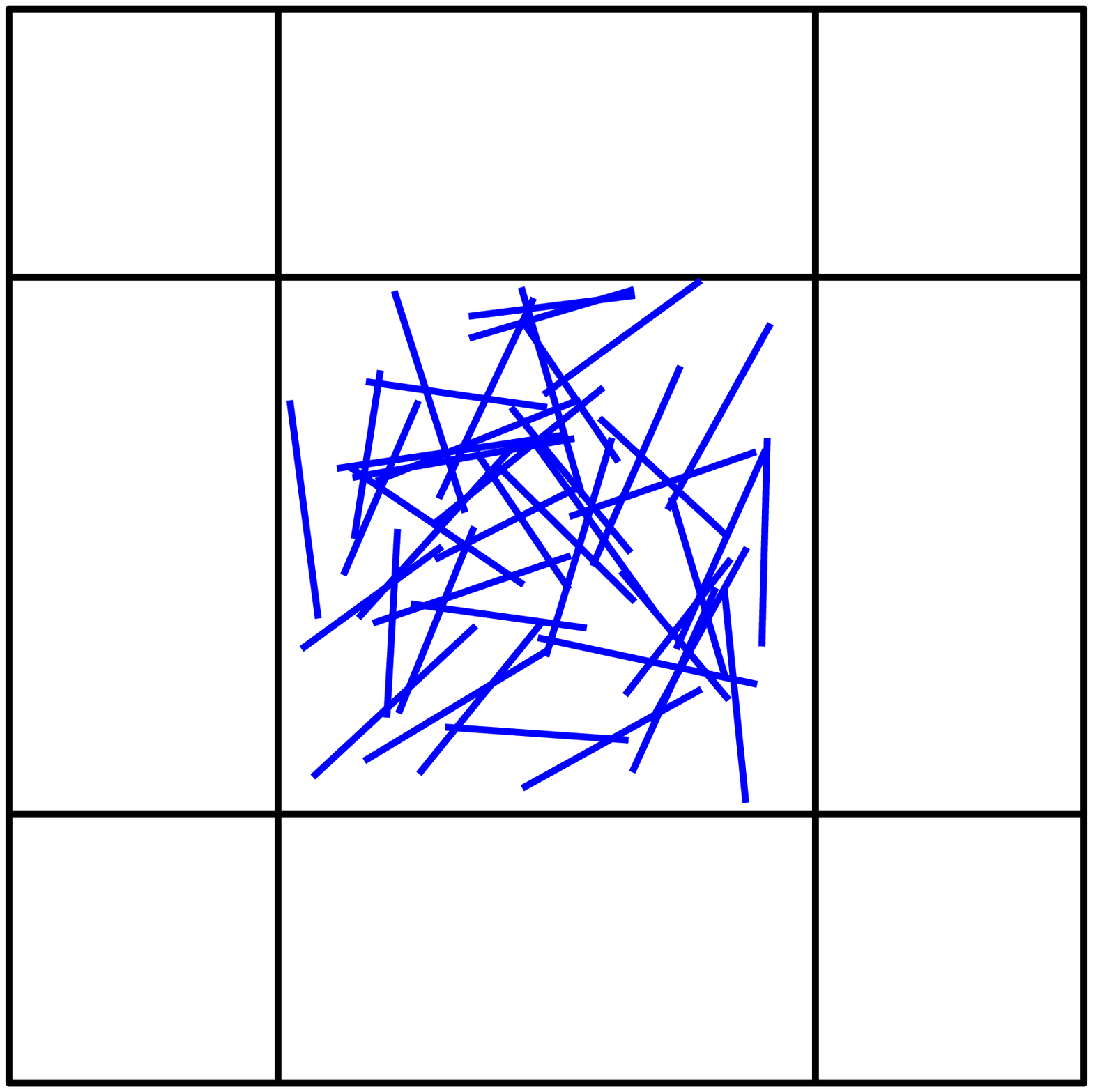}
\end{centering}
 \caption{Two basic schemes for dividing a set of galaxy pairs into sub-samples: For each sub-region of the survey, there will be galaxy pairs crossing from that region into another (upper panel, green and red). In the galaxy-scheme $\boldsymbol \xi^{\alpha}$ is computed by considering only pairs that completely lay within the sub-region $\alpha$ (lower right panel).  In the pair-jackknife scheme (lower left panel) half of the pairs that cross from $\alpha$ to another region (drawn in green) are taken into account for computing $\boldsymbol \xi^{\alpha}$ while only the other half (red) is discarded.}
\label{fig:jack_schemes}
\end{figure}

\subsection{Galaxy pairs crossing between sub-samples}
\label{sec:cross_pairs}

A problem specific to the internal covariance estimation for two-point correlation functions is the question of what to do with pairs of galaxies where each galaxy lies in a different sub-region of the survey. 

In fact, the pieces of information in a cosmic shear survey are not the individual galaxy shapes but the pairs of galaxy shapes. If the pairs crossing between sub-regions are completely ignored when computing the sub-measurements $\hat{\xi}^\alpha$, then one is re-sampling a data set that has \emph{less} information than the total measurement of $\xi_\pm$ and hence a larger variance. Note, that this does not only influence the shape-noise part of the covariance but also the cosmic variance part. The reason is that galaxies at the edge of a sub-region contribute less terms to the correlation function measurement than galaxies in the center of the sub-region (c.f. figure \ref{fig:effective_area}), i.e. the area of the sub-patch is not uniformly probed by the galaxy pairs and the measured shear correlations are dominated by the inner part of the patch. In contrast to the correlation of sub-samples discussed before, this increases the cosmic variance between the sub-samples and can bias the covariance estimate high - especially on large angular scales.

This effect can in principle be resolved by re-sampling the set of pairs (instead of the set of galaxy shapes), i.e. by defining the sub-measurement $\hat{\boldsymbol \xi}^{\alpha}$ as
\begin{flalign}
\label{eq:pair_jackknife}
\hat \xi_\pm^\alpha(\theta) = \phantom{\hspace{7cm}} \nonumber \\
\nonumber \\
\frac{\sum_{\mathrm{pairs}\ \mathrm{in}\ \alpha} (\epsilon_t^i \epsilon_t^j \pm \epsilon_\times^i \epsilon_\times^j) + \sum_{\mathrm{half}\ \mathrm{of}\ \mathrm{cross}\ \mathrm{pairs}} (\epsilon_t^i \epsilon_t^j \pm \epsilon_\times^i \epsilon_\times^j)}{N_{\mathrm{pairs}}}\ . \nonumber \\
\end{flalign}
How this resampling of galaxy pairs can be done is illustrated in figure \ref{fig:jack_schemes}. Especially one has to make sure that each galaxy pair enters exactly one of the $\hat{\boldsymbol \xi}^{\alpha}$. We call this procedure the \emph{pair-scheme} while we will call the standard procedure of considering only the individual galaxies in sub-region $\alpha$ when computing $\hat{\boldsymbol \xi}^{\alpha}$ as \emph{galaxy-scheme}.

\begin{figure*}
\begin{centering}
\includegraphics[width=0.95\textwidth]{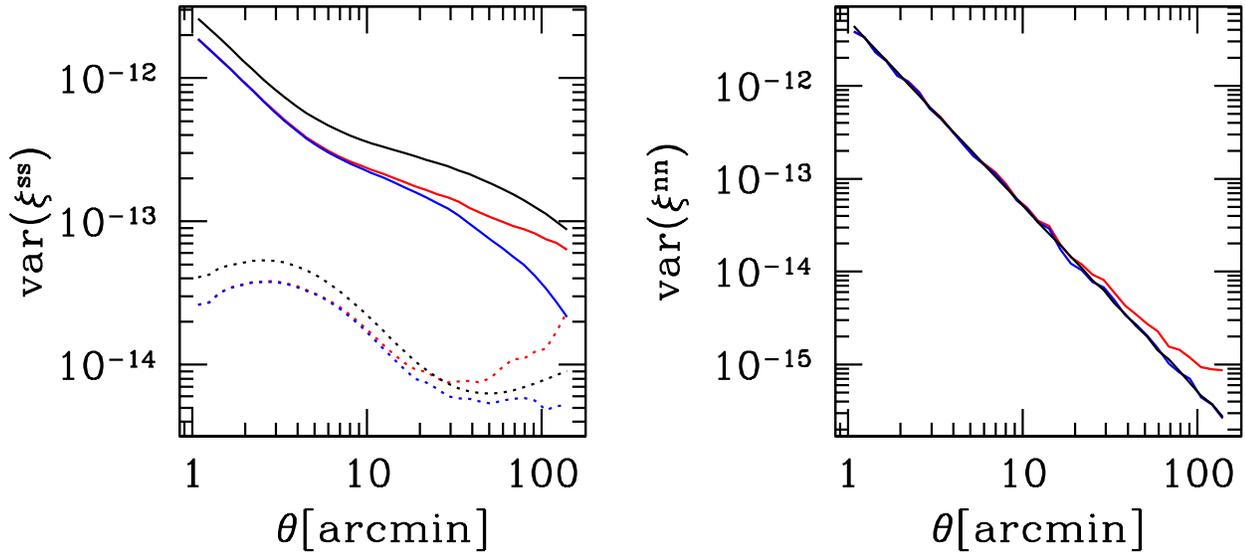}
\end{centering}
 \caption{Different variance estimates using the sub-sample covariance estimator and $400$ sub-samples. \underline{Left:} Variance estimates for $\hat{\boldsymbol \xi}_+$ (solid lines) and $\hat{\boldsymbol \xi}_-$ (dotted lines) in a mock catalog without shape noise that is otherwise following setup I. The red lines show the galaxy-scheme estimate (c.f. section \ref{sec:cross_pairs}), the blue lines show the pair-scheme estimate and the black lines show the log-normal input model. \underline{Right:} Sub-sample estimates of the variance of $\hat{\boldsymbol \xi}_+$ in a mock catalog that only consists of shape noise and has the same area and density as in setup I. It is only in this situation (and in the pair-scheme) that internal estimation of the covariance yields unbiased results.}
\label{fig:cov_components}
\end{figure*}

In figure \ref{fig:cov_components} we demonstrate this effect along with the effect of correlated sub-samples that was discussed before. The left-hand panel shows sub-sample estimates of the variance of $\hat\xi_\pm$ in a simulated survey (corresponding to setup I in table \ref{tab:setup}) where the shape noise was put to zero and where $400$ sub-samples were used. Both the variance of $\hat\xi_+$ and $\hat\xi_-$ are severely underestimated on small scales, which is due to the correlation of sub-samples. At large angular scales, the galaxy-scheme yields systematically higher value for the variance than the pair-scheme and at least for $\hat\xi_-$ it can even overestimate the variance. This is due to the missing cross-pairs in the re-sampling.

For the right-hand panel of figure \ref{fig:cov_components} we have generated a catalog of pure shape noise ($\sigma_\epsilon, A, n_{\mathrm{gal}}$ as in setup I). This is the only situation where the assumption of uncorrelated sub-samples is valid. You can see that in this case the pair-scheme is able to estimate the variance without bias. The galaxy-scheme overestimates the variance for the reasons explained before. A downside of the pair-scheme is that the shear signals in the sub-measurements $\hat{\boldsymbol \xi}^{\alpha}$ become even more correlated, as can also be seen from the left-hand panel of figure \ref{fig:cov_components}.

\subsection{Jackknife}

Another method of covariance estimation that \citet{Norberg2009} investigate is the \emph{delete-one-jackknife}. Instead of estimating the covariance of the measurements $\boldsymbol \xi^\alpha$ and rescaling it to the size of the whole survey the jackknife is considering the measurements
\begin{equation}
\label{eq:gaussianity_sum4}
\hat \xi_\pm^{*\alpha}(\theta) = \frac{\sum_{\lbrace i, j\ \mathrm{not}\ \mathrm{in}\ \alpha\rbrace} (\epsilon_t^i \epsilon_t^j \pm \epsilon_\times^i \epsilon_\times^j)\cdot \Delta_\theta (|\boldsymbol \theta_i - \boldsymbol \theta_j|)}{\sum_{\lbrace i, j\ \mathrm{not}\ \mathrm{in}\ \alpha\rbrace} \Delta_\theta (|\boldsymbol \theta_i - \boldsymbol \theta_j|)}\ ,
\end{equation}
i.e. the \emph{jackknife-sample} $\alpha$ is generated by cutting out the subregion $\alpha$ and measuring the correlation functions in the rest of the survey. The jackknife estimate for the covariance matrix is then given by \citep{Efron1982, Norberg2009}
\begin{equation}
\label{eq:def_jack}
\hat C_{\mathrm{jack}} = \frac{N-1}{N} \sum_{\alpha = 1}^N (\boldsymbol \xi^{*\alpha}- \boldsymbol{\bar\xi}^*)^T (\boldsymbol \xi^{*\alpha}- \boldsymbol{\bar\xi}^*)\ ,
\end{equation}
where $\boldsymbol{\bar\xi}^*$ is the mean of all jackknife measurements.

If we again assume that all subregions have the same galaxy density and that the correlation functions are measured on scales much smaller than the sub-region size then $\boldsymbol \xi^{*\alpha}$ is approximately given by
\begin{equation}
\boldsymbol \xi^{*\alpha} \approx \frac{1}{N-1} \sum_{\beta \neq \alpha} \hat{\boldsymbol \xi}^{\beta}\ .
\end{equation}
From this it also follows that
\begin{eqnarray}
\boldsymbol \xi^{*\alpha}-\bar{\boldsymbol \xi}^* &\approx & \frac{1}{N-1} \sum_{\beta \neq \alpha} \hat{\boldsymbol \xi}^{\beta} - \frac{1}{N}\sum_{\beta} \boldsymbol \xi^{*\beta} \nonumber \\
&=&\frac{N\cdot \bar{\boldsymbol \xi} - \hat{\boldsymbol \xi}^{\alpha}}{N-1} - \frac{1}{(N-1)\cdot N}\sum_{\beta}\sum_{\gamma\neq \beta} \hat{\boldsymbol \xi}^{\gamma} \nonumber \\
&=& \frac{N\cdot \bar{\boldsymbol \xi} - \hat{\boldsymbol \xi}^{\alpha}}{N-1} - \frac{N-1}{(N-1)\cdot N}\sum_{\gamma}\hat{\boldsymbol \xi}^{\gamma} \nonumber \\
&=& \frac{N\cdot \bar{\boldsymbol \xi} - \hat{\boldsymbol \xi}^{\alpha}}{N-1} - \bar{\boldsymbol \xi} \nonumber \\
&=& \frac{\bar{\boldsymbol \xi} - \hat{\boldsymbol \xi}^{\alpha}}{N-1}\ .
\end{eqnarray}
Inserting this into the definition of $\hat C_{\mathrm{jack}}$ gives exactly the subsample covariance $\hat C_{\mathrm{SC}} $, i.e. on small angular scales the two methods are approximately equivalent\footnote{This is no general statement on the jackknife method. It holds only in our particular situation.}.

In jackknife estimation one can in principle also differentiate between a pair scheme and a galaxy scheme. Using eq. \ref{eq:gaussianity_sum4} for $\boldsymbol \xi^{*\alpha}$ corresponds to the galaxy scheme. This is equivalent to disregarding all pairs in the top panel of figure \ref{fig:jack_schemes} when computing $\boldsymbol \xi^{*\alpha}$. The pair-scheme is given by disregarding all pairs in the lower left panel of figure \ref{fig:jack_schemes} when computing $\boldsymbol \xi^{*\alpha}$. In the pair scheme jackknife and sub-sample covariance become exactly equivalent when (assuming that each sub-patch has the same number of galaxies).

\subsection{Bootstrap Covariance}

The so called \emph{block bootstrap} estimator of the covariance also divides the data into sub-samples. If the data is split into $N$ sub-regions, then a number of $N_{\mathrm{boot}}$ bootstrap re-samplings of the data are generated by randomly drawing with replacement $N$ of the sub-samples and combining then into one re-sampled data set \citep{Norberg2009, Nordmann2007, Loh2008, Efron1982}. If the correlation function measured in the re-sampled data $i$ ($i = 1,\ \dots\ ,\ N_{\mathrm{boot}}$) is called ${\boldsymbol \xi}^{\mathrm{boot} ,i}$, then the bootstrap estimate of the covariance is given by
\begin{equation}
\hat C_{\mathrm{boot}} = \frac{1}{N_{\mathrm{boot}}-1} \sum_{i = 1}^{N_{\mathrm{boot}}} ({\boldsymbol \xi}^{\mathrm{boot} ,i}- \boldsymbol{\bar\xi}^{\mathrm{boot}})^T\ ({\boldsymbol \xi}^{\mathrm{boot} ,i} - \boldsymbol{\bar\xi}^{\mathrm{boot}})\ ,
\end{equation}
where $\boldsymbol{\bar\xi}^{\mathrm{boot}}$ is now the mean of all ${\boldsymbol \xi}^{\mathrm{boot} ,i}$.

Again, the question arises of whether one should consider the single galaxies or the galaxy pairs as the actual data (cf. section \ref{sec:cross_pairs}). In what we will call \emph{galaxy-bootstrap} one simply adds a copy of all galaxies in a sub-region $\alpha$ to the re-sampled data set $i$ each time the sub-region $\alpha$ gets drawn.

In the \emph{pair-bootstrap} one adds all pairs associated to sub-region $\alpha$ to the list of pairs that is used to compute ${\boldsymbol \xi}^{\mathrm{boot} ,i}$. The difference between the two bootstrap schemes is mainly the following: if the sub-region $\alpha$ gets drawn $n$ times, then each pair in $\alpha$ gets a weight of $n$ in the pair-scheme and a weight of $n^2$ in the galaxy-scheme.

Note that the pair-bootstrap is very similar to what \citet{Loh2008} describes as \emph{marked point bootstrap}, the only difference being, that we chose to split pairs between sub-regions evenly among these regions. 

We will see in section \ref{sec:jackknife_results} that the galaxy-bootstrap severely overestimates the covariance. The other covariance estimators perform very similar to each other and suffer in similar ways from the systematics explained in subsection \ref{sec:subsample_correlation} and \ref{sec:cross_pairs}.

\subsection{Stability and Inversion of the Covariance Estimate}
\label{sec:inversion}

All effects that bias the internal covariance estimate can in principle be minimized by dividing the data into very large sub-regions. This decreases both the correlation of the different sub-regions and the influence of pairs crossing between sub-regions. However, this also decreases the possible number of re-samplings and hence increases the variance of the covariance estimator itself.

In order to derive constraints on the number of re-samplings let us assume that we are in the limit were the correlations between sub-regions are small. Small here means that
\begin{equation}
\langle \Delta\hat{\boldsymbol \xi}_i^\alpha\Delta\hat{\boldsymbol \xi}_j^\beta \rangle \ll \langle \Delta\hat{\boldsymbol \xi}_i^\alpha\Delta\hat{\boldsymbol \xi}_j^\alpha \rangle\ , \ \mathrm{for}\ \alpha \neq \beta\ .
\end{equation}
As explained before, this is the only limit in which internal covariance estimation is valid. In this limit the sub-sample covariance is just a rescaling of the sample covariance of independent realizations of the sub-regions. Hence - in the limit considered here and under the assumption that the data vector behaves Gaussian -  the sub-sample covariance estimates are distributed according to a Wishart distribution (cf. \citealt{Taylor2012}).  Also, the pair-jackknife is almost equivalent to the pair-version of the sub-sample covariance, i.e. to equation \ref{eq:subsample_estimator} when $\hat{\boldsymbol \xi}^{\alpha}$ is computed with equation \ref{eq:pair_jackknife}. Hence, also the pair-jackknife estimates should approximately follow a Wishart distribution.

The most important consequence of this is that the inverse of the covariance matrix estimate will be a biased estimate of the true inverse covariance matrix, and the bias is approximately given by \citep{Hartlap2007, Taylor2012}:
\begin{equation}
\langle \hat C_{\mathrm{SC}}^{-1} \rangle \approx \frac{N-1}{N-d-2} C_{\mathrm{true}}^{-1}\ ,
\end{equation}
where $N$ is the number of sub-regions and $d$ is the number of data points in $\hat{\boldsymbol \xi}$. Especially, this factor has to be accounted for when computing the $\chi^2$ statistic, eq. \ref{eq:figure_of_merit}, i.e. it has an influence on the constraints derived on cosmological parameters when using internal covariance estimation.

\textcolor{blue}{
\citet{Taylor2012} also give constraints on $N$ with respect to $d$ when a certain accuracy in the final parameter constraints is demanded.\footnote{However, they are ignoring the impact that the variance in the inverted covariance estimate has on parameter constraints, which is investigated by \citet{Taylor2014}.} We take their criterion,
\begin{equation}
\label{eq:parameter_accuracy}
N \overset{!}{>} \frac{2}{\epsilon^2} + (d+4)\ ,
\end{equation}
where $\epsilon$ is the required fractional accuracy on parameter constraints, as a guideline also for internal covariance estimation. This is however under the assumption of an exact Wishart distribution, i.e. that the data vector is Gaussian and that the sub-regions are large enough to not cause systematic biases in the covariance estimate. Demanding a fractional accuracy of $\epsilon = 0.2$ for the parameter constraints, this yields a necessary number of $N > 54 + d$ re-sampling. Below this number there is no chance for internal covariance estimation to yield parameter constraints that are accurate to more that $20\%$.}

\section[]{Testing internal Covariance Estimators on simulated cosmic Shear Surveys}
\label{sec:jackknife_results}

\begin{figure}
\center{
\includegraphics[width=0.45\textwidth]{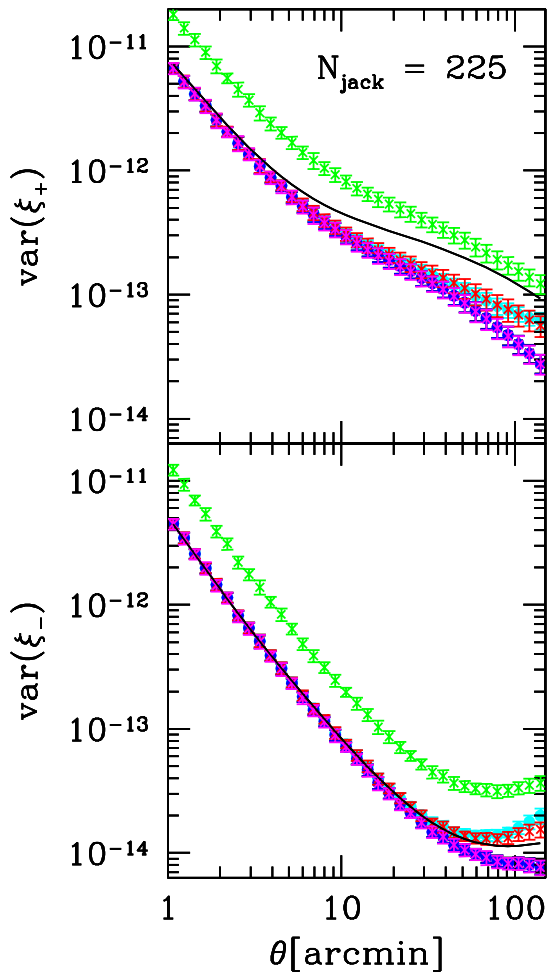}
}
 \caption{A comparison of the different internal estimation schemes when splitting the survey into $N=225$ sub-regions. Green: galaxy-bootstrap, purple: pair-bootstrap, red: galaxy-jackknife, blue: pair-jackknife and cyan: sub-sample covariance compared to the analytical covariance (black line). We show the sub-sample covariance only in the galaxy-scheme because in the pair-scheme it is almost identical to jackknife and bootstrap. As explained in section \ref{sec:jackknife_schemes}, at large angular scales the different treatment of galaxy pairs crossing between sub-region leads to an overestimation of the variance by the galaxy-scheme and an underestimation of the variance by the pair-scheme.}
 \label{fig:method_comparison}
\end{figure}

We will now use the simulations described in section \ref{sec:log-normal_simulations} to test the performance of internal covariance estimators. First, we will use setup I (cf. \ref{tab:setup}) corresponding to a rather deep survey. We carry out $50$ independent realizations of this survey. In each survey we measure the correlation functions in the range and binning that was explained in section \ref{sec:log-normal_simulations}. We then estimate the covariance of the measured correlation functions using the different internal estimation schemes that were introduced in section \ref{sec:jackknife_schemes}. Throughout this section - except for subsection \ref{sec:des_sv} - we consider the log-normal model that was explained in section \ref{sec:covariance_models} as the \emph{true} covariance of the simulated surveys. This is justified by the fact that our results don't change if we instead use the sample covariance of $1000$ independent realisations that were presented in section \ref{sec:log-normal_simulations} (cf. appendix \ref{sec:appendix_off_diagonal}, figure \ref{fig:constraints_empirical_covariance}).

In figure \ref{fig:method_comparison} we compare the sub-sample, jackknife and bootstrap estimates of the diagonal elements of the covariance matrix (both in the galaxy- and pair-scheme) when splitting the survey into $N=225$ sub-regions. The most impressive finding is, that in the galaxy-scheme the bootstrap severely overestimates the variance. This is in agreement with the findings of \citet{Norberg2009} for galaxy clustering correlation functions. The duplication of whole sub-volumes of galaxies creates bootstrap samples that are in fact unrealistic, i.e. these bootstrap samples contain regions with no sources at all and on the other hand regions with a very high source density. Each original galaxy pair gets weighted by a factor of $n^2$ when the corresponding region is drawn $n$ times. This puts a very high weight on small sub-areas of the bootstrap sample and creates an unphysically high variance between the bootstrap samples. 

In the pair-scheme however, all three internal estimators perform almost identical. This is not surprising, because in that scheme the bootstrap is just an approximation to the sub-sample covariance and sub-sample and jackknife covariance are almost identical in the pair-scheme. As explained in section \ref{sec:cross_pairs}, in the galaxy-jackknife scheme the two effects of correlated sub-regions and false re-sampling of pairs partly cancel each other. Hence the galaxy-jackknife comes closest to the true variance at large scales. The performance of the sub-sample covariance (in the galaxy-scheme) only slightly differs from that.

\begin{figure*}
\center{
\includegraphics[width=0.9\textwidth, height=1.3\textwidth]{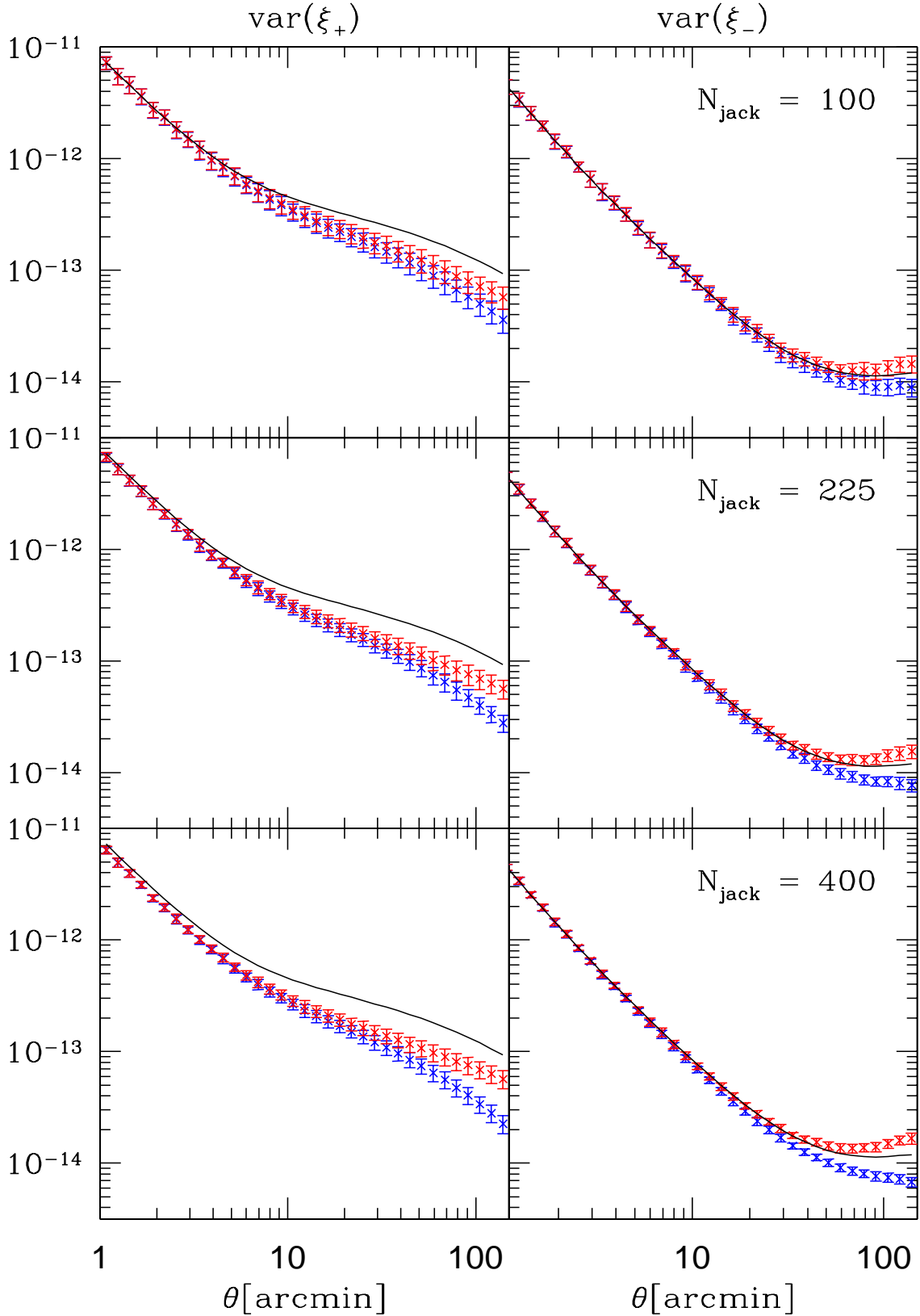}
}
 \caption{Mean values of $50$ jackknife estimates of the variance of $\xi_+$ (left) and $\xi_-$ (right).
 Galaxy-jackknife was used for the red points while pair-jackknife was used for the blue points and the errorbars show the sample standard deviation of the single estimate (as estimated from the $50$ jackknife matrices).
 The black line corresponds to the log-normal input model of the simulations.}
 \label{fig:jack_estimates}
\end{figure*}

Because of the strong similarity between the different estimator we will restrict the following analyses to the pair-jackknife and the galaxy-jackknife. We now investigate the influence of sub-region size on internal covariance estimation. Hence we split the surveys into $3$ different numbers of sub-regions: $10^2$, $15^2$ and $20^2$ corresponding to sub-region areas of approximately $7.0\times 7.0\ \deg^2$, $4.67\times 4.67\ \deg^2$ and $3.5\times 3.5\ \deg^2$. In figure \ref{fig:jack_estimates} we compare the mean value of the $50$ jackknife estimates of the variance of $\hat\xi_\pm$ (the diagonal elements of the covariance matrix) to the true underlying log-normal model. A comparison of the off-diagonal behaviour of the jackknife estimates to that of the input-covariance can be found in appendix \ref{sec:appendix_off_diagonal}. The errorbars in figure \ref{fig:jack_estimates} represent the standard deviation of the $50$ jackknife estimates, i.e. they illustrate the noise of the internal estimators. You can see in this figure the biases in the jackknife estimates that we explained in the previous section. For $\xi_+$, both jackknife schemes underestimate the variance. At large scales, this is in the galaxy-jackknife scheme partly compensated by the false re-sampling of galaxy pairs. For $\xi_-$, the pair-jackknife underestimates the variance while the galaxy-jackknife overestimates it. $\xi_-$ is a much more local measure in the sense that the different sub regions are less correlated in $\hat\xi_-$ and that the covariance matrix is much more dominated by the shape noise contributions. Hence, the severe systematic underestimation of the variance that can be seen for $\xi_+$ does not appear as strongly for $\xi_-$.

When increasing the number of sub-regions for the jackknife estimators, the noise in the variance estimates becomes smaller but the deviations from the true variance also become stronger. This is because for smaller sub-regions the estimated $\hat{\boldsymbol \xi}^\alpha $ become more correlated and because there will be more galaxy pairs crossing from one sub-region to another.

\subsection{Constraints on cosmological Parameters}

We will now take the $50$ simulations as mock observations and try to constrain the dark matter density parameter $\Omega_m$ and the power spectrum normalization $\sigma_8$. To do so we sample the $\Omega_m$-$\sigma_8$ plane on a fine grid while keeping the other cosmological parameters fixed. Following a Bayesian approach we take the probability density in the parameter space to be proportional to the likelihood,
\begin{equation}
\label{eq:likelihood}
p(\boldsymbol \pi) \sim \mathcal{L}(\boldsymbol \pi) \sim \exp\left( -\frac{1}{2} \chi^2[\boldsymbol \pi] \right) \ , 
\end{equation}
where we assume our data vector $\hat{\boldsymbol \xi}$ to be Gaussian such that
\begin{equation}
\chi^2[\boldsymbol \pi] = (\hat{\boldsymbol \xi} - \boldsymbol \xi[\boldsymbol \pi])^T C^{-1}\ (\hat{\boldsymbol \xi} - \boldsymbol \xi[\boldsymbol \pi])\ .
\end{equation}
Here $\boldsymbol \xi[\boldsymbol \pi]$ are our model predictions for $\langle \hat{\boldsymbol \xi} \rangle $ which we again compute with the \textsc{nicaea} package. We are assuming a prior of $\Omega_m \in \left[ 0.1 , 0.4\right]$ and $\sigma_8 \in \left[ 0.8 , 1.1\right]$, which is well centered around our input cosmology. For $C$ we will either insert the log-normal model covariance or the jackknife estimates of the covariance. We will de-bias the inverse of the latter in the way explained in section \ref{sec:inversion}. Note that the reasoning in section \ref{sec:inversion} is in principle only valid for the pair-jackknife. And also for the pair-jackknife it is only valid in the case of almost uncorrelated sub-regions. We will nevertheless carry out the de-biasing in the same way for both jackknife schemes. Furthermore, we will also ignore the variance of the inverted covariance estimate \citep{Taylor2014}, as explained in the end of section \ref{sec:jackknife_schemes}. Our data vector $\hat{\boldsymbol \xi}$ will be either $\hat{\boldsymbol \xi}_+$ or $\hat{\boldsymbol \xi}_-$ or the joint data vector of both correlation functions, in which case we will also take into account the cross covariance between the two.

For each mock observation $\hat{\boldsymbol \xi}$ and for each available covariance matrix we use equation \ref{eq:likelihood} to compute marginalised $1\sigma$ constraints on $\Omega_m$ and $\sigma_8$, i.e. we consider the marginalised probability densities
\begin{eqnarray}
p_\Omega(\Omega_m) &=& \int \mathrm{d}\sigma_8\ p(\Omega_m , \sigma_8) \nonumber \\
p_\sigma(\sigma_8) &=& \int \mathrm{d}\Omega_m\ p(\Omega_m , \sigma_8)
\end{eqnarray}
and we define $1\sigma$ confidence interval to be that interval around the best fit parameter value which encloses $\sim 68\%$ of the probability and which has equal values of the probability density at each interval boundary\footnote{Without the last statement the definition of the $1\sigma$ confidence interval would be ambiguous.}.

Because of the strong degeneracy between $\Omega_m$ and $\sigma_8$ \citep{Kilbinger1, Kilbinger2}, even little uncertainties in the modelling of $\boldsymbol \xi[\boldsymbol \pi]$\footnote{In our modelling we are for example not considering the finite bin width in our measurement of $\hat{\boldsymbol \xi}$.} or in our simulations could shift the best-fit values of the parameters along the degeneracy. Fortunately, this does not affect our analysis because we only have to compare the constraints derived from the jackknife covariance estimates to the constraints obtained from the true (log-normal) covariance matrix. Furthermore, our results don't change noticeably, if instead of the log-normal covariance matrix we use the sample covariance estimated from 1000 simulations (c.f. section \ref{sec:log-normal_simulations}). Hence in any case, our analysis provides a fair test of internal covariance estimators.

\begin{figure}
\begin{centering}
\includegraphics[width=0.49\textwidth]{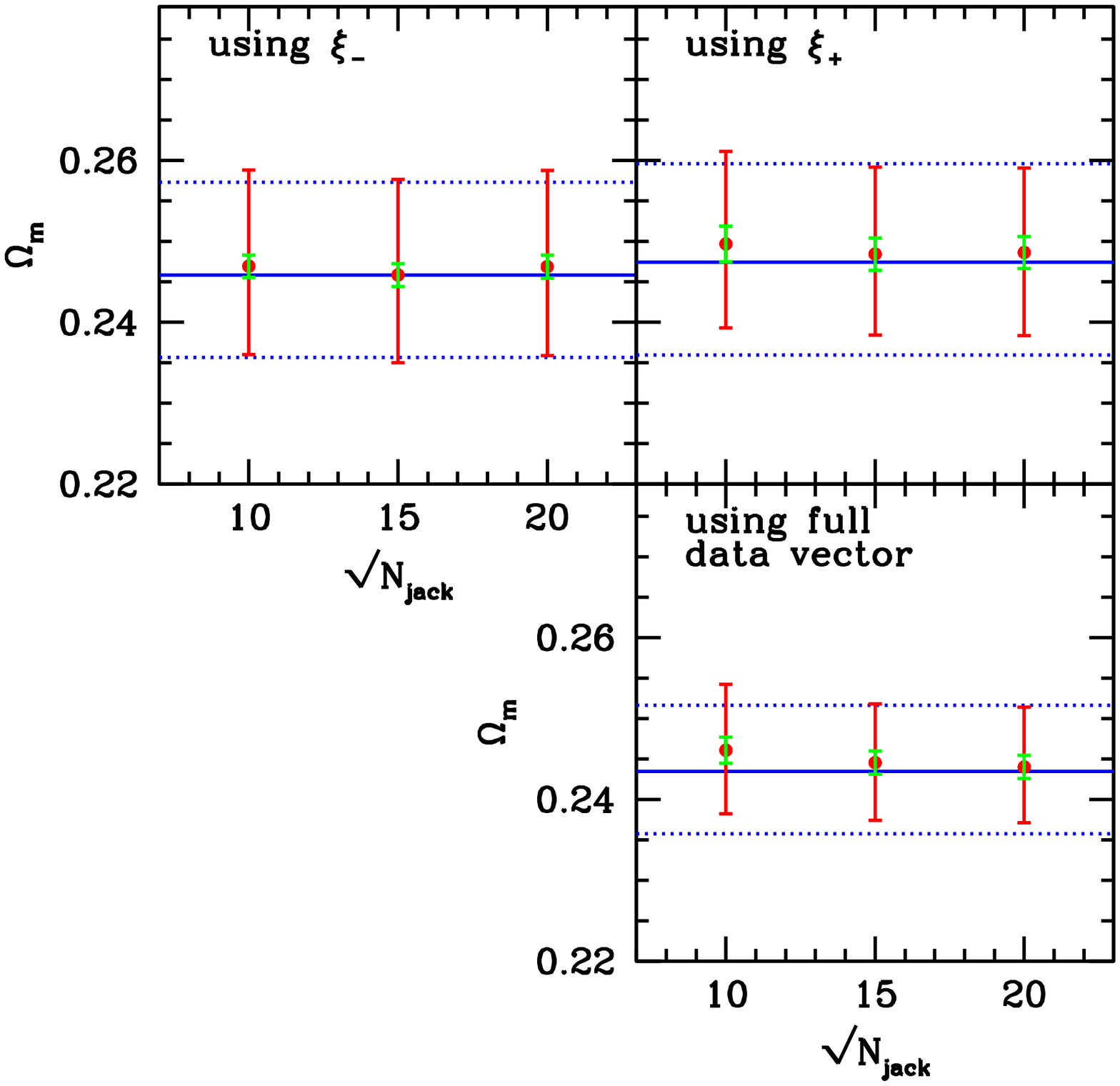}

\includegraphics[width=0.49\textwidth]{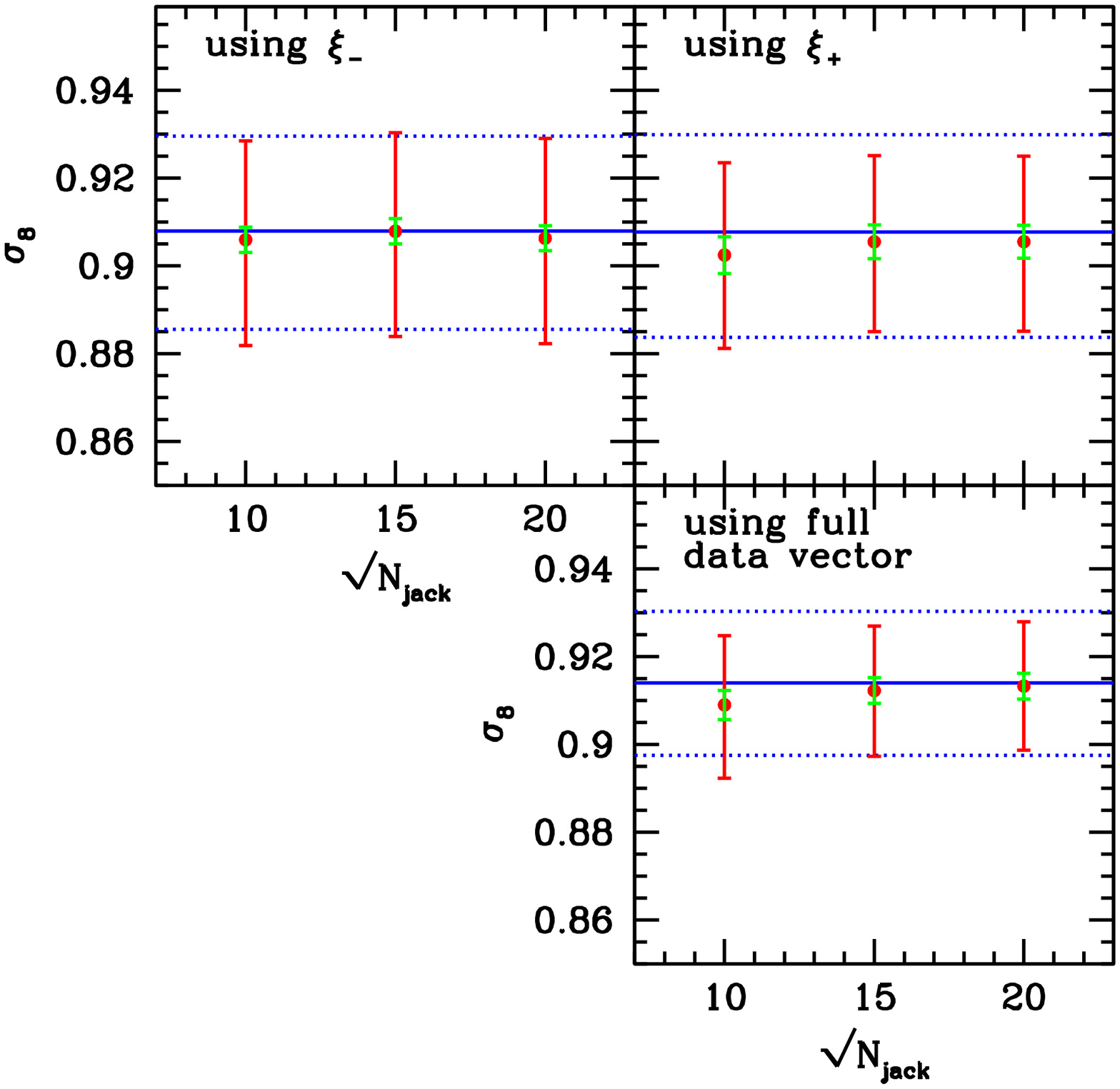}
\end{centering}
 \caption{Mean $1\sigma$ constraints on $\Omega_m$ and $\sigma_8$ using galaxy-jackknife (red errorbars). The green errorbars show the standard deviation of the mean best-fit values (i.e. the standard deviation of the best fit values divided by $\sqrt{50}$). The blue lines indicate the constraints that are obtained when the true covariance is used in each mock catalog. Note that the error bars are very symmetric. For surveys as big as our simulations the constraining power becomes large enough to turn the - usually \emph{banana} shaped - degeneracy between $\Omega_m$ and $\sigma_8$ into almost elliptical contours in the parameter plane (c.f. appendix \ref{sec:appendix})}
 \label{fig:constraints}
\end{figure}

In figure \ref{fig:constraints} we show the mean values of the upper and lower boundaries on $\Omega_m$ and $\sigma_8$ as well as their mean best fit value for different numbers of jackknife re-samplings (red points and errorbars). The mean is taken with respect to all $50$ confidence intervals we computed from the $50$ mock observations. We also compare the jackknife constraints to those we get when using the true covariance matrix (blue lines). These figures only show the results for the galaxy-jackknife, which in the situation considered here yields the best agreement with the true covariance.

\begin{figure}
\begin{centering}
\includegraphics[width=0.49\textwidth]{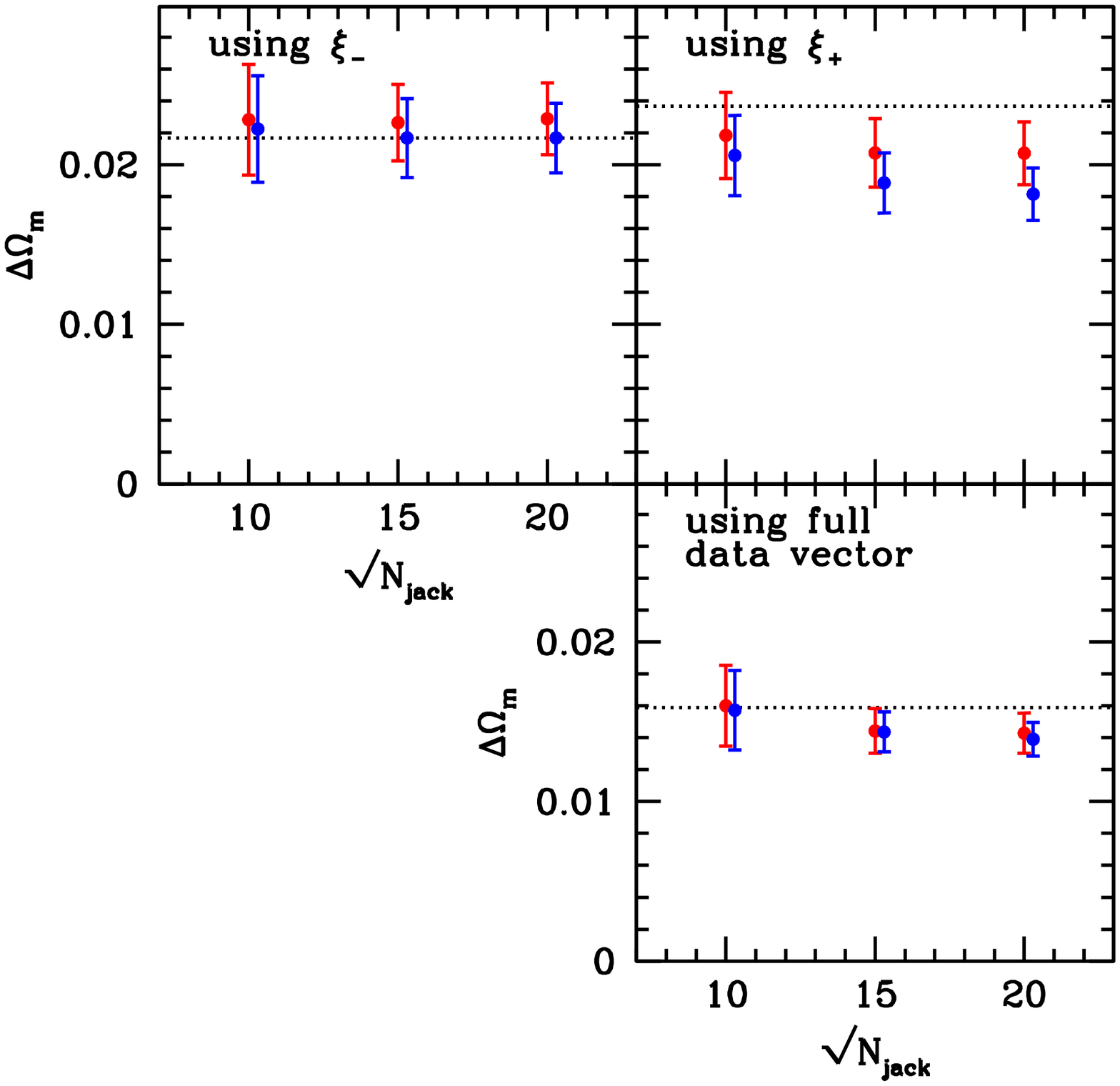}

\includegraphics[width=0.49\textwidth]{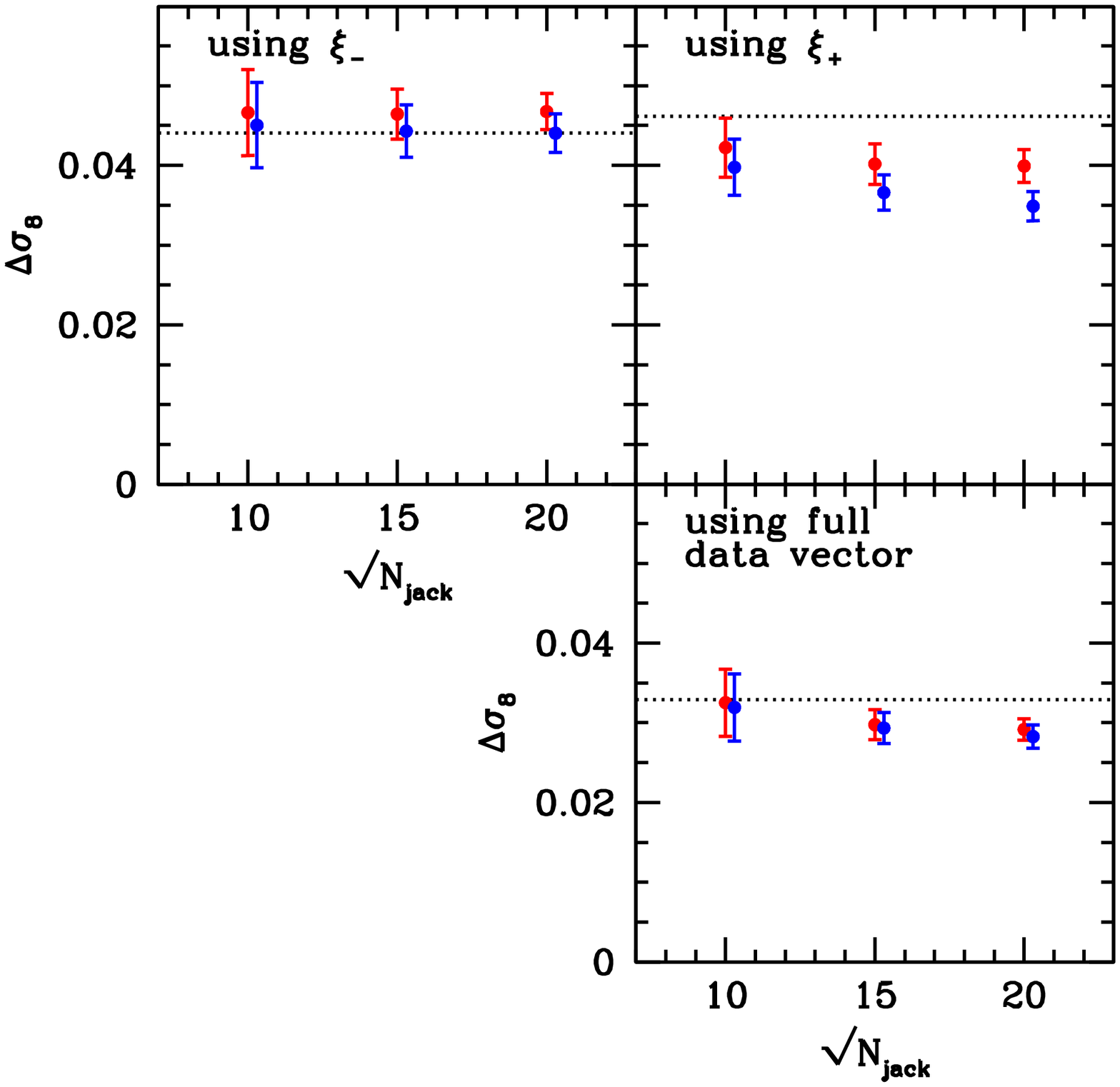}
\end{centering}
 \caption{Mean width of the $1\sigma$ uncertainty on $\Omega_m$ and $\sigma_8$ using pair-jackknife (blue) and galaxy-jackknife (red). The errorbars show the standard deviation the $50$ estimated confidence intervals. The black dotted line indicates the mean width of the confidence intervals when the true covariance is used in each mock catalog.}
 \label{fig:interval_width}
\end{figure}

We compare galaxy-jackknife and pair-jackknife in figure \ref{fig:interval_width}. Here we show the mean width of the confidence intervals obtained with galaxy-jackknife, pair-jackknife and the true covariance matrix. For $\xi_-$, the width of the confidence intervals agrees well with the confidence intervals obtained from the true covariance matrix. This is because the covariance matrix of $\xi_-$ is dominated by its shape noise component, which is very accurately captured by jackknife. In fact, even for the pair-scheme and even for $400$ jackknife re-samplings the width of the confidence intervals from $\xi_-$ alone is not underestimated. This seems to contradict figure \ref{fig:jack_estimates}, where the pair-scheme systematically underestimates the covariance. One reason for this is probably, that the variance in the inverted covariance estimate increases parameter uncertainties \citep{Taylor2014}. Note especially, that this is not the same effect as the de-biasing in eqn. \ref{sec:inversion}. For $\xi_+$, the strong underestimation of the covariance matrix by jackknife also leads to an underestimation of the uncertainties on $\Omega_m$ and $\sigma_8$. Again one can see that the variance in the width of the confidence intervals (the errorbars in figure \ref{fig:interval_width}) becomes smaller, when more jackknife re-samplings are used. In turn, this increases the overall underestimation of the uncertainties. If both correlation functions are combined and $225$ re-samplings are used, the parameter uncertainties are underestimated by $\sim 10\%$.

\begin{figure}
\begin{centering}
\includegraphics[width=0.49\textwidth]{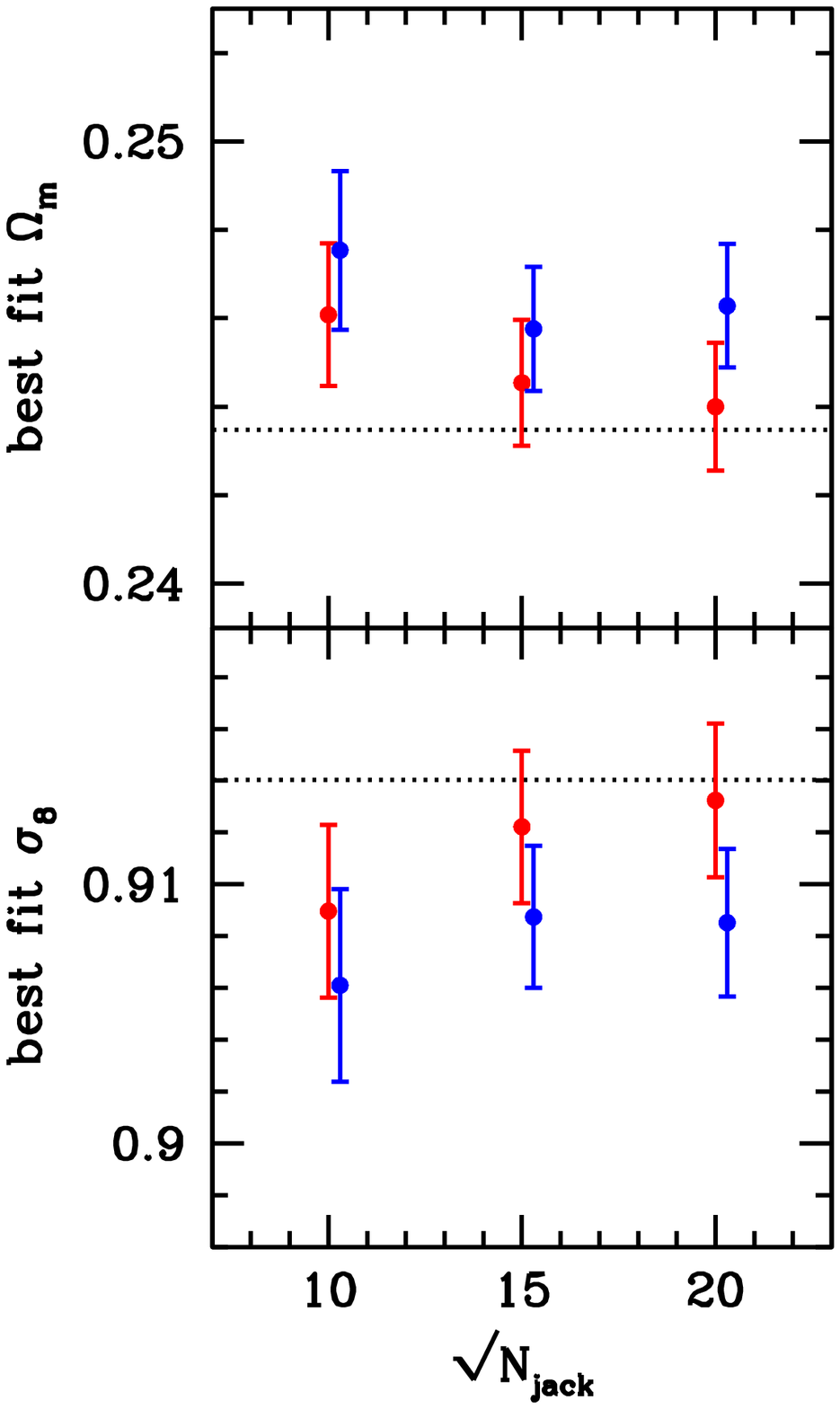}
\end{centering}
 \caption{Mean best fit values of $\Omega_m$ and $\sigma_8$ using pair-jackknife (blue) and galaxy-jackknife (red). The errorbars show the standard deviation of the mean, as estimated from the $50$ best fit values. The black dotted line indicates the mean best fit value when the true covariance is used in each mock catalog.}
 \label{fig:best_fits}
\end{figure}

We have not shown results from the pair-jackknife estimates in figure \ref{fig:constraints}, but the best fit values of $\Omega_m$ and $\sigma_8$ agree very well between the two jackknife schemes (i.e. within the green errorbars in figure \ref{fig:constraints}), if only $\hat{\boldsymbol \xi}_+$ or $\hat{\boldsymbol \xi}_-$ are used to constrain the parameters. In figure \ref{fig:best_fits} we compare the pair-jackknife and galaxy-jackknife best fit values when using the \emph{full} data vector. Here the pair-jackknife seems to yield a stronger bias of the best fit values with respect to the true covariance.

The above results indicate that internal covariance estimation can reproduce the constraints on parameters from the true covariance quite well, especially when the galaxy-jackknife scheme is used. However, these results are not generalizable. In general, internal estimation of the covariance works best if the covariance matrix is shape noise dominated. Hence, the answer to what is the best estimation scheme and how well it can reproduce the true errorbars on cosmological parameters depends on the depth of the considered survey. A shallower survey not only has a smaller source density and hence a bigger shape noise. It also has a smaller convergence power spectrum which in turn reduces the cosmic variance part of the covariance.

The procedure we presented above to investigate the performance of internal covariance estimators thus has to be re-run for each survey under consideration. One can consider the log-normal model as a good model for the true covariance of our simulations for mock catalogs with an area of $\gtrsim 1000 \deg^2$ and a simple, connected geometry. For smaller surveys the finite-area-effect should not be ignored \citep{Sato2011, Kilbinger1}. However, these surveys can be simulated fast enough with our public code to generate a large sample of independent realisations of the mock data which provides a good sample covariance estimate of the true covariance matrix. This estimate can then be compared to an ensemble of internal covariance estimates as we have done it above.

\subsection{Matching the procedure to DES science verification and year 5 Data}
\label{sec:des_sv}

We will now present an application of our method. Our attempt is to determine the performance of internal covariance estimation for
\begin{itemize}
\item setup IIa: Dark Energy Survey science verification data (DES-SV)
\item setup IIb: DES year five data (DES-Y5) assuming a low source density
\item setup IIc: DES year five data assuming a high source density.
\end{itemize}
For the area, shape noise, source density and source redshift distribution cf. table \ref{tab:setup} and section \ref{sec:setup}. A source density of $10\ {\mathrm{arcmin}}^{-2}$ is forecasted for the final DES data while a density of $\sim 6\ {\mathrm{arcmin}}^{-2}$ roughly corresponds to the current status of DES science verification data. Note also, that we are using a mask similar to the footprint of DES-SV to simulate mock shape catalogs for setup IIa. Setups IIb and IIc are simply simulated to be square shaped.

We adjust our data vector to that used by \citet{Becker2015}, i.e. for both $\xi_+$ and $\xi_-$ we now use 15 logarithmic bins ranging from $\theta = 2$ arcmin to $\theta = 300$ arcmin. We will cut the survey into 100 sub-regions for setup IIa. Note that this way our biggest angular scales by far exceed the diameter of our subregions which is $\sim 45$ arcmin. Hence, this can be considered an on-the-edge test of internal covariance estimators. A good tool to define sub-regions in an irregular survey geometry is the \emph{kmeans} algorithm\footnote{implemented by Erin Sheldon for python, www.github.com/esheldon/kmeans\_radec}. For setups IIb and IIc we decide to split the survey into $225$ sub-regions which corresponds to a diameter of $\sim 4.7$ arcmin. This should give a more stable estimate of the covariance while still yielding much larger sub-regions than in setup IIa.

\begin{figure*}
\begin{centering}
\includegraphics[width=0.32\textwidth]{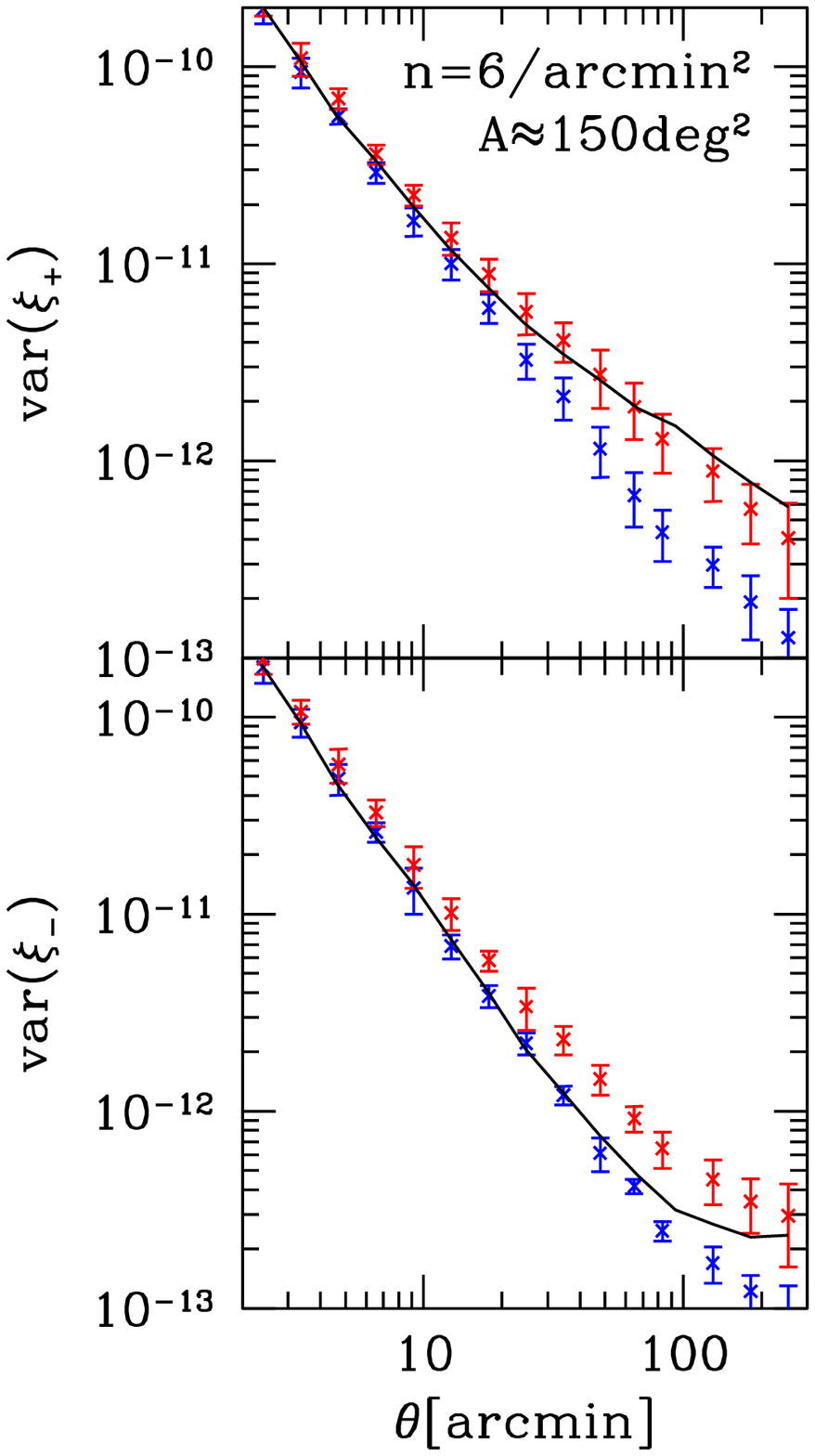}\includegraphics[width=0.32\textwidth]{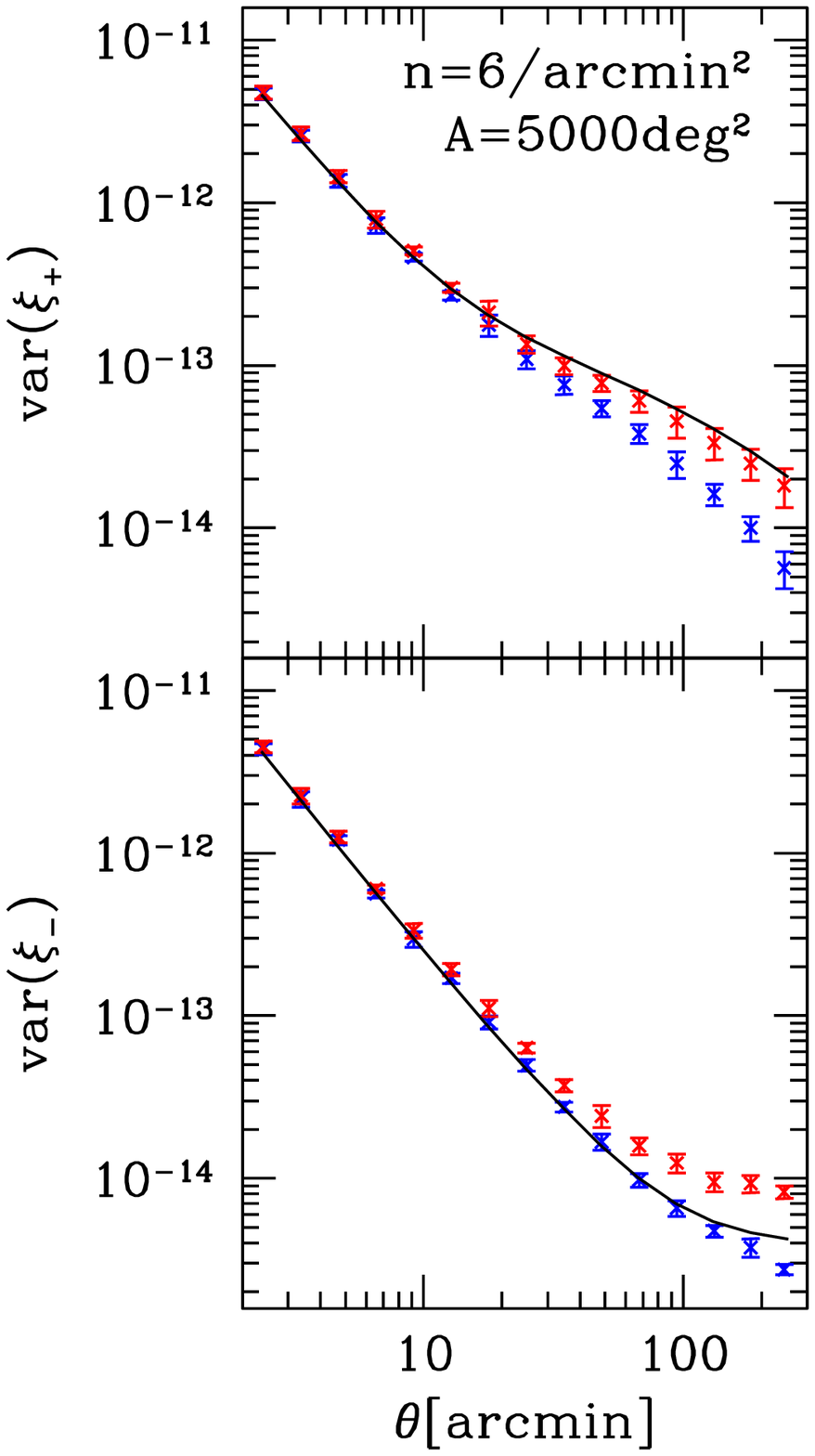}\includegraphics[width=0.32\textwidth]{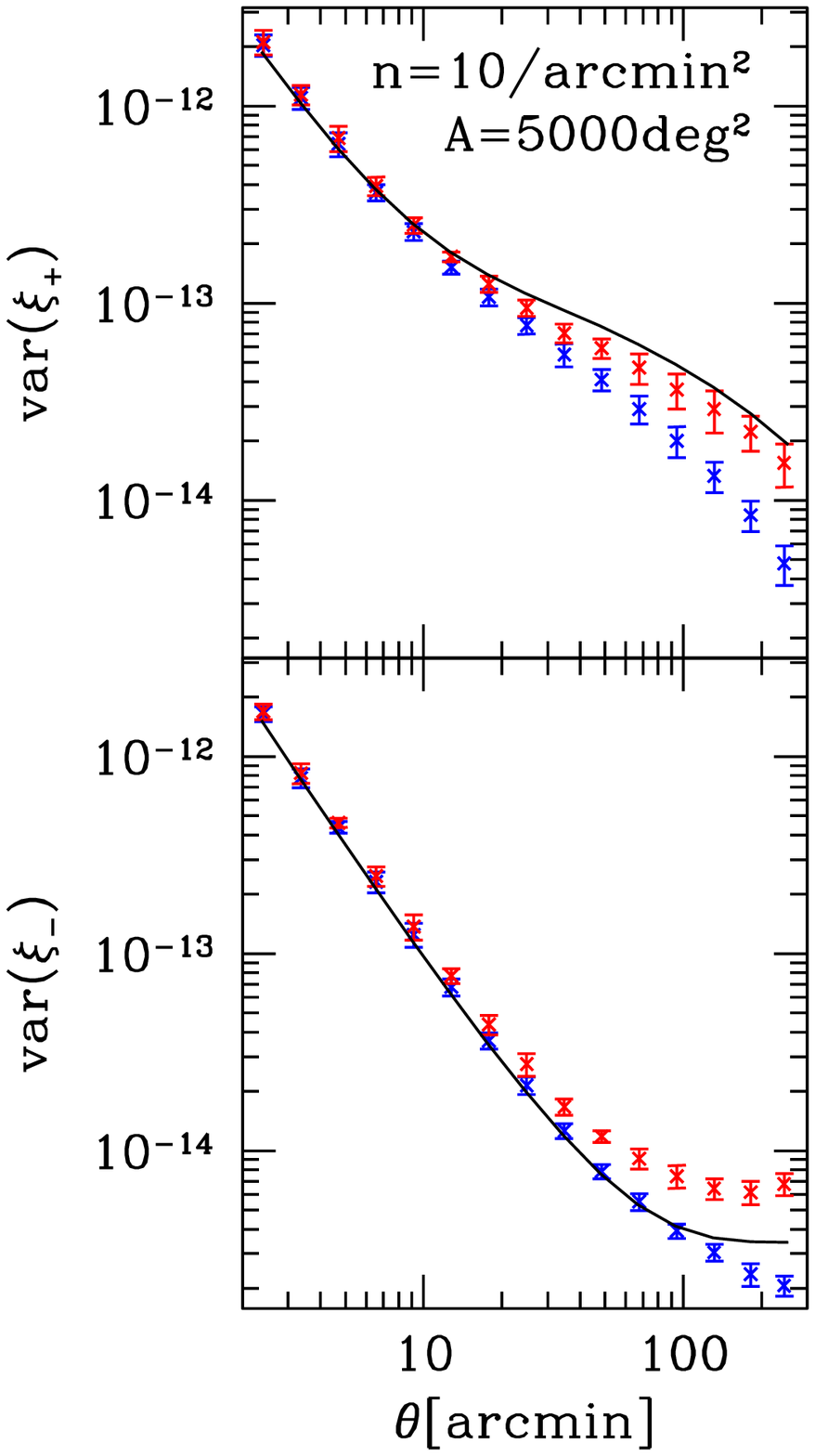}
\end{centering}
 \caption{Variance estimates for DES-SV like data (left), DES-Y5 like data with a low density (middle) and with a high density (right). Red dots show the galaxy-jackknife estimates and blue dots the pair-jackknife estimates. For the Y5 case the lognormal model together with eqn. \ref{eq:finite_bin_covariance} was taken as a reference covariance (black lines) while for the SV case we estimated the true covariance from $1000$ independent realizations of the mock data in order to account for the finite-area-effect. The errorbars indicate the standard deviation of the single estimates as obtained from $10$ independent measurements.}
 \label{fig:sv_variance}
\end{figure*}
In figure \ref{fig:sv_variance} we compare the internal variance estimates to the true covariance. The latter is taken to be the log-normal model for the Y5 simulations and a sample variance computed from $1000$ independent realisations for the SV simulations. Because of the fewer number of bins we are now using the procedure described in section \ref{sec:finite_bins} to compute the log-normal covariance matrix. As you can see, for $\hat\xi_-$ the pair-jackknife now becomes the best estimator of the variance. For $\hat\xi_+$ the situation is similar to what we have seen before, i.e. both schemes mostly underestimate the variance and the galaxy-jackknife is overall closer to the true variance. Hence, judging from figure \ref{fig:sv_variance} we conclude that galaxy-jackknife should be used in order to not under estimate the true uncertainties in the data vector. However, these statements only hold for the diagonal elements of the covariance matrix. A convenient way to compare the complete covariance estimates is to derive likelihood contours from them in the desired parameter space.

\textcolor{red}{
We carry out a likelihood analysis in the $\Omega_m$-$\sigma_8$ plane for the $10$ simulations that have a Y5-like area and a source density of $6\ {\mathrm{arcmin}}^{-2}$ which is the highest density currently achieved in DES science verification data \citep{Becker2015}. In figure \ref{fig:likelihood_contours} we show the likelihood contours obtained from one of the simulations when using galaxy-jackknife, pair-jackknife and the log-normal model for the covariance matrix. The contours were obtained from Monte-Carlo-Markov-Chains (150.000 steps) using the COSMOLIKE package by \citet{Eifler2014}. We present the likelihood contours from the other $9$ independent simulations in appendix \ref{sec:appendix}. As expected, jackknife estimation underestimates the uncertainties. The input cosmology lies within the $1$-$\sigma$ contour in 6 of 10 simulation, when the log-normal covariance is used. It lies within the $1$-$\sigma$ contour in 5 of 10 simulation, when the covariance is estimated with jackknife (either scheme). 
}

\textcolor{red}{
In table \ref{tab:ratios_2D} we show the average ratio of the volume in the $\Omega_m$-$\sigma_8$ plane enclosed by the $1\sigma$- and $2\sigma$-contours when using jackknife to that when using the true covariance matrix. Since the $1\sigma$- and $2\sigma$-ellipses obtained from jackknife and from the true covariance lie well on top of each other, this ratio can be considered as the fraction of the true uncertainties that is recovered by the jackknife covariance matrices. You can see from table \ref{tab:ratios_2D} that the volume inside contours of constant likelihood in the $\Omega_m$-$\sigma_8$ plane estimated with galaxy-jackknife is on average $\gtrsim 85\%$ of the true volume while the volume estimated with pair-jackknife recovers only $\gtrsim 70\%$ of the true volume. This agrees with the impression (from figures \ref{fig:contours1} and \ref{fig:contours2}) that the contours obtained with galaxy-jackknife match better to the contours obtained from the true covariance. Note also, that the ellipses obtained from pair-jackknife have in some cases a strong off-set along the degeneracy between $\Omega_m$ and $\sigma_8$ compared to the true covariance and the galaxy-jackknife estimates. This is probably because pair-jackknife strongly underestimates the variance of $\hat\xi_\pm$ at large angular scales, which causes even small fluctuations at these scales to shift the contours considerably.
}

\textcolor{red}{
Finally, we want to see how well jackknife matrices recover the uncertainties perpendicular to the degeneracy between $\Omega_m$ and $\sigma_8$. To do so, we consider the parameter combination
\begin{equation}
\Sigma_8 := \frac{\sigma_8}{0.9} \left(\frac{\Omega_m}{0.25}\right)^{0.5}\ .
\end{equation}
Contours of constant $\Sigma_8$ are roughly parallel to the degeneracy that can be seen in figures \ref{fig:likelihood_contours}, \ref{fig:contours1} and \ref{fig:contours2}. For each of our $10$ realisations we bin our MCMC's in $\Sigma_8$ to estimate its probability density. Table \ref{tab:ratios_1D} displays the average ratio of the $1\sigma$ and $2\sigma$ uncertainties obtained from jackknife to the uncertainties obtained from the true covariance. This time, we find that galaxy-jackknife on average yields $\sim 90 \%$ of the true uncertainties while pair-jackknife yields $\sim 85\%$. Hence, when the degeneracy between $\Omega_m$ and $\sigma_8$ is broken by other probes (such as the power spectrum of temperature fluctuation in the cosmic microwave background) the performance of jackknife covariance matrices slightly improves.
}

\textcolor{red}{
Judging from the above numbers and from the contours in appendix \ref{sec:appendix} we deem that $\gtrsim 85\%$ of the true uncertainties on $\Omega_m$ and $\sigma_8$ in a 2D cosmic shear analysis can be recovered without the use of large suits of N-body simulations or covariance models.  When other probes like the CMB are used to break the degeneracy between the two parameters, the performance of jackknife even increases, because the deviations from the true covariance mostly take place along the direction of degeneracy between $\Omega_m$ and $\sigma_8$.
}

\section{Conclusions}
\label{sec:conclusions}

We have explored the performance of internal covariance estimation for cosmic shear 2-pt. correlation functions. We devised two different jackknife schemes and explained in detail when these schemes underestimate the true covariance and when overestimation takes place. Furthermore, we explained why the sub-sample covariance and the pair-bootstrap covariance yield results that are very similar to jackknife estimation of the covariance matrix. Based on the pair-jackknife scheme we have argued that the Anderson-Hartlap-Kaufman \citep{Kaufman,Hartlap2007} de-biasing factor should also be applied when inverting jackknife covariance matrices. Based on empirical findings we also recommend the use of this factor for the galaxy-jackknife scheme.

\textcolor{blue}{
We have demonstrated our findings in an exemplary study using log-normal simulations of the convergence field and the corresponding shear field. We found the performance of all internal covariance estimators - except for the bootstrapping of galaxies - to be very similar. For the investigated cases, jackknife covariance matrices could provide accurate uncertainties on cosmological parameters as compared to the true covariance matrix of our simulations. Our conclusions regarding the two possible re-sampling schemes are the following:}
\begin{itemize}
\item \textcolor{blue}{galaxy-bootstrap severely overestimates the covariance, which is in agreement with the finding of \citet{Norberg2009} for galaxy clustering correlation functions.}
\item \textcolor{blue}{from $\xi_-$ alone, the pair-jackknife scheme reconstructs the parameter constraints most faithfully (cf. figure \ref{fig:constraints}). }

\item \textcolor{blue}{from $\xi_+$ alone and when combining the two correlation functions, we find that the parameter constraints are best reconstructed by the galaxy-jackknife.}
\end{itemize}
\textcolor{blue}{The performance of the galaxy-scheme turns out to be better in most situations, because two systematic errors (cf. sections \ref{sec:subsample_correlation} and \ref{sec:cross_pairs}) cancel each other partly in the that scheme. The pair-jackknife suffers from only one of these systematics and hence always yields lower (absolute) values for the covariance than the galaxy-jackknife and always underestimates the (absolute) values of the true covariance matrix.}

\begin{figure}
\includegraphics[width=0.5\textwidth]{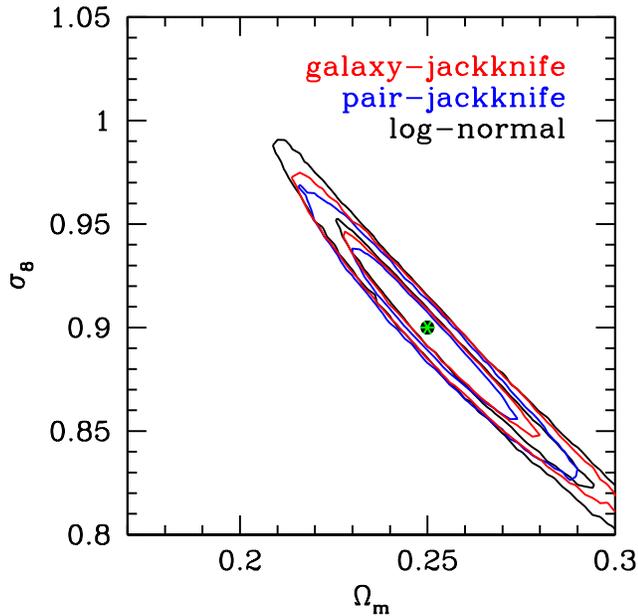}
\caption{$1$-$\sigma$ and $2$-$\sigma$ contours in the $\Omega_m$-$\sigma_8$ plane obtained from the two jackknife schemes (read and blue) and the true covariance (log-normal covariance, black) and using the combined data vector $(\hat{\xi}_+, \hat{\xi}_-)$. The input cosmology lies within the $1$-$\sigma$ contour in 6 of 10 simulation, when the log-normal covariance is used. It lies within the $1$-$\sigma$ contour in 5 of 10 simulation, when the covariance is estimated with jackknife (either scheme). In appendix \ref{sec:appendix} we show the contours obtained from the other simulations. The underestimation of the uncertainties by jackknife mainly takes place along the direction of the degeneracy between $\Omega_m$ and $\sigma_8$.}
 \label{fig:likelihood_contours}
\end{figure}

Our results can not be generalized to arbitrary surveys, i.e. our paper rather demonstrates a general method to find a good covariance estimation scheme for any particular survey. In making our simulation code public we provide our readers with a tool to re-do the presented analyses for their desired set-up. As an application example we tested jackknife estimation of the covariance for a 2D cosmic shear analysis of the Dark Energy Survey. We found that for the complete, 5-year DES survey internal covariance estimators can provide reliable parameter constraints in a 2D cosmic shear analysis. We recommend a scheme of $\sim 15\times 15$ jackknife re-samplings to yield a stable covariance matrix. Judging from figures \ref{fig:likelihood_contours}, \ref{fig:contours1} and \ref{fig:contours2}, we find as before that the likelihood contours in the $\Omega_m$-$\sigma_8$ plane are best reconstructed by the galaxy-jackknife scheme, if both correlation functions $\xi_+$ and $\xi_-$ are combined. \textcolor{red}{This way, on average $\gtrsim 85\%$ of the true uncertainties are captured by the internally estimated covariance matrix. If the degeneracy between $\Omega_m$ and $\sigma_8$ is broken, this value increases to $\sim 90\%$. Hence, up to $\sim 90\%$ of the true uncertainties in a 2D cosmic shear analysis can be provided from internally estimated covariance matrices.}

\FloatBarrier

\newpage

\begin{table}
\begin{tabular}{ c | c | c } 
\/ & galaxy-jackknife & pair-jackknife \\ 
\hline
$V_{1\sigma , \mathrm{jack}}/V_{1\sigma , \mathrm{true}}$ & $0.86\pm 0.08$ & $0.72\pm 0.09$ \\ 
$V_{2\sigma , \mathrm{jack}}/V_{2\sigma , \mathrm{true}}$ & $0.87\pm 0.08$ & $0.74\pm 0.09$ \\ 
\end{tabular}
\caption{Ratio of the volume within the $1\sigma$ and $2\sigma$ contours in the $\Omega_m-\sigma_8$ plane obtained from jackknife and true covariance (setup IIb). The errors are given by the standard deviation of a sample of $10$ independent simulations. The combined data vector of $\xi_+$ and $\xi_-$ was used.}
 \label{tab:ratios_2D}
\end{table}

\begin{table}
\begin{tabular}{ c || c | c } 
\/ & galaxy-jackknife & pair-jackknife \\ 
\hline
$\Delta\Sigma_{8\ 1\sigma , \mathrm{jack}}/\Delta\Sigma_{8\ 1\sigma , \mathrm{true}}$ & $0.91\pm 0.08$ & $0.86\pm 0.10$ \\ 
$\Delta\Sigma_{8\ 2\sigma , \mathrm{jack}}/\Delta\Sigma_{8\ 2\sigma , \mathrm{true}}$ & $0.90\pm 0.08$ & $0.85\pm 0.09$ \\ 
\end{tabular}
\caption{Ratio of the $1\sigma$ and $2\sigma$ uncertainties on $\Sigma_8 \sim \sigma_8 \Omega_m^{0.5}$ obtained from jackknife and true covariance (setup IIb). The errors are given by the standard deviation of a sample of $10$ independent simulations. Again, the combined data vector of $\xi_+$ and $\xi_-$ was used.}
 \label{tab:ratios_1D}
\end{table}

\section*{Acknowledgments}

This work was supported by SFB-Transregio 33 `The Dark Universe'  by  the  Deutsche  Forschungsgemeinschaft  (DFG). We also acknowledge the support by the DFG Cluster of Excellence "Origin and Structure of the Universe". The simulations have been carried out on the computing facilities of the Computational Center for Particle and Astrophysics (C2PAP). Part of the research was carried out at the Jet Propulsion Laboratory, California Institute of Technology, under a contract with the National Aeronautics and Space Administration.

This paper has gone through internal review by the DES collaboration. We thank David Bacon, Gary Berstein, Stefan Hilbert, Klaus Honscheid, Benjamin Joachimi and Bhuvnesh Jain for very helpful comments and discussions during the review process.

Funding for the DES Projects has been provided by the U.S. Department of Energy, the U.S. National Science Foundation, the Ministry of Science and Education of Spain, 
the Science and Technology Facilities Council of the United Kingdom, the Higher Education Funding Council for England, the National Center for Supercomputing 
Applications at the University of Illinois at Urbana-Champaign, the Kavli Institute of Cosmological Physics at the University of Chicago, 
the Center for Cosmology and Astro-Particle Physics at the Ohio State University,
the Mitchell Institute for Fundamental Physics and Astronomy at Texas A\&M University, Financiadora de Estudos e Projetos, 
Funda{\c c}{\~a}o Carlos Chagas Filho de Amparo {\`a} Pesquisa do Estado do Rio de Janeiro, Conselho Nacional de Desenvolvimento Cient{\'i}fico e Tecnol{\'o}gico and 
the Minist{\'e}rio da Ci{\^e}ncia, Tecnologia e Inova{\c c}{\~a}o, the Deutsche Forschungsgemeinschaft and the Collaborating Institutions in the Dark Energy Survey. 
The DES data management system is supported by the National Science Foundation under Grant Number AST-1138766.

The Collaborating Institutions are Argonne National Laboratory, the University of California at Santa Cruz, the University of Cambridge, Centro de Investigaciones En{\'e}rgeticas, 
Medioambientales y Tecnol{\'o}gicas-Madrid, the University of Chicago, University College London, the DES-Brazil Consortium, the University of Edinburgh, 
the Eidgen{\"o}ssische Technische Hochschule (ETH) Z{\"u}rich, 
Fermi National Accelerator Laboratory, the University of Illinois at Urbana-Champaign, the Institut de Ci{\`e}ncies de l'Espai (IEEC/CSIC), 
the Institut de F{\'i}sica d'Altes Energies, Lawrence Berkeley National Laboratory, the Ludwig-Maximilians Universit{\"a}t M{\"u}nchen and the associated Excellence Cluster Universe, 
the University of Michigan, the National Optical Astronomy Observatory, the University of Nottingham, The Ohio State University, the University of Pennsylvania, the University of Portsmouth, 
SLAC National Accelerator Laboratory, Stanford University, the University of Sussex, and Texas A\&M University.

The DES participants from Spanish institutions are partially supported by MINECO under grants AYA2012-39559, ESP2013-48274, FPA2013-47986, and Centro de Excelencia Severo Ochoa SEV-2012-0234.
Research leading to these results has received funding from the European Research Council under the European Union’s Seventh Framework Programme (FP7/2007-2013) including ERC grant agreements 
 240672, 291329, and 306478.

%BIBLIOGRAPHY
\bibliographystyle{mn2e}
\bibliography{cosmic_shear_covariance_estimation}

%\bsp

\newpage

\appendix 

\section{Correlation Matrices and Constraints from empirical Covariance}
\label{sec:appendix_off_diagonal}

\begin{figure}
\begin{centering}
\includegraphics[width=0.49\textwidth]{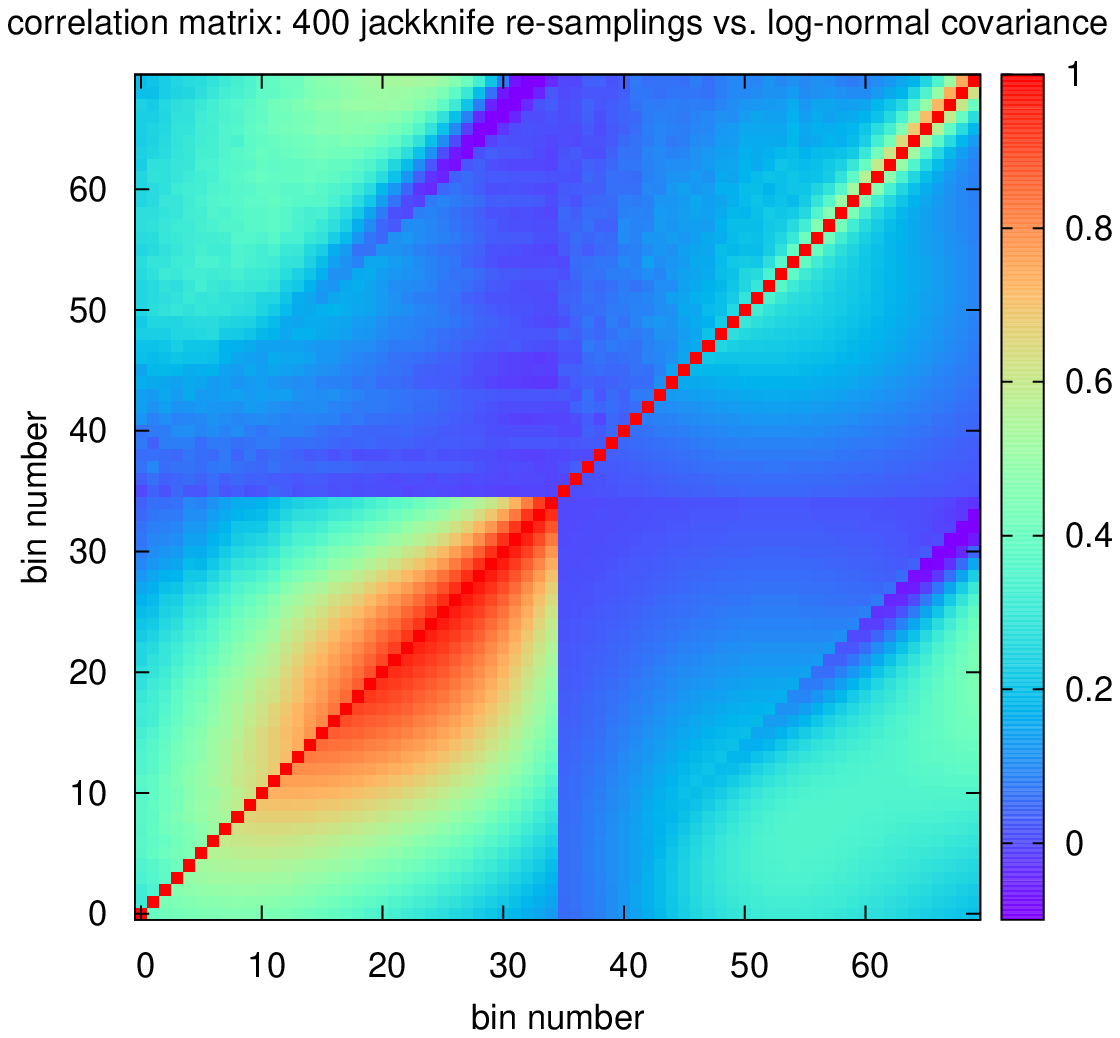}

\includegraphics[width=0.49\textwidth]{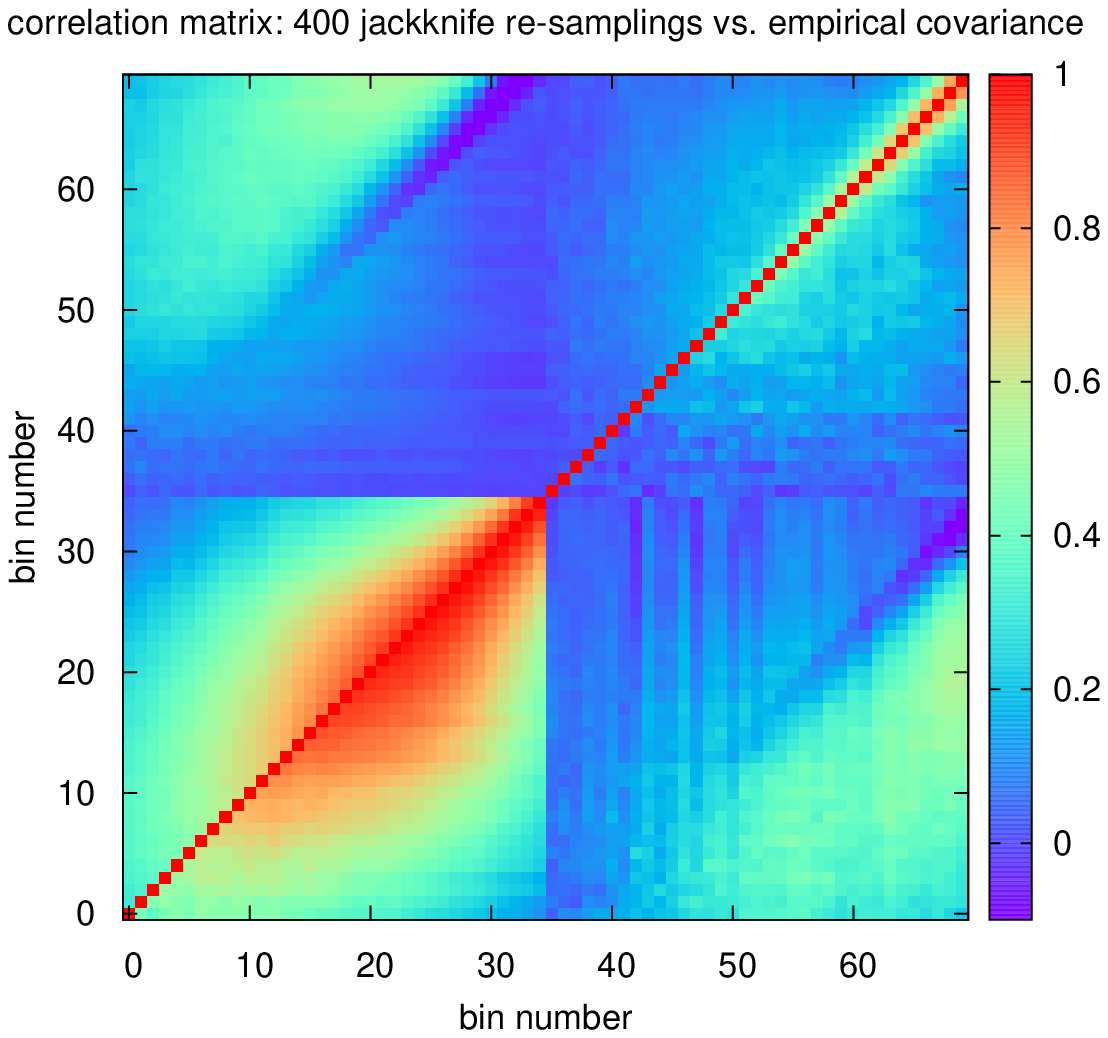}
\end{centering}
 \caption{Correlation matrix of $\hat{\xi}_\pm$. Bins $0$ to $34$ belong to $\hat{\xi}_+$ and bins $35$ to $69$ belong to $\hat{\xi}_-$. For each 2-pt function the bins range from $1'$ to $150'$, starting on the lower left corner. Top panel: the lower right half of the plot displays the correlation coefficients of $\hat{\xi}_\pm$ obtained from the log-normal model and the upper left halt displays the correlation coefficients obtained from the average jackknife covariance estimate (using 400 re-samplings). Bottom panel: same plot, but this time the empirical covariance from $1000$ realizations of setup I. Note that we are averaging over $50$ different jackknife estimates of the covariance matrix here, which makes the empirical covariance matrix seem noisier than the jackknife covariance.}
 \label{fig:correlation_matrix}
\end{figure}

To see how jackknife estimates of the covariance matrix capture the cross-correlations between different bins of $\hat{xi}_\pm$ we are looking at the \emph{correlation matrix}. This matrix is given in terms of the covariance matrix elements $\mathrm{Cov}_{ij}$ as
\begin{equation}
\label{eq:correlation_matrix}
\mathrm{Corr}_{ij} = \frac{\mathrm{Cov}_{ij}}{\sqrt{\mathrm{Cov}_{ii}\mathrm{Cov}_{jj}}}\ .
\end{equation}
In the top panel of figure \ref{fig:correlation_matrix} we compare the correlation matrix obtained from the log-normal model covariance to the correlation matrix obtained from averaging all $50$ jackknife estimates of the covariance matrix that were presented in section \ref{sec:jackknife_results}, i.e. using setup I from table \ref{tab:setup} (we show here the jackknife with $400$ re-samplings which divided the survey into the smallest sub-regions). The lower left corner shows the auto correlations of $\hat{\xi}_+$ and the upper right corner shows the auto correlations of $\hat{\xi}_-$. The upper left and lower right corners show the cross-correlations between the two correlation functions. Furthermore, the lower right half of the plot shows the correlations obtained from the log-normal model and the upper left half shows the correlations obtained from the average jackknife covariance estimate. Each column and row of pixels represents one angular bin and the bins range from $1'$ to $150'$, starting on the lower left corner.

The top panel of figure \ref{fig:correlation_matrix} indicates, that jackknife is able to capture the general structure of the correlation matrix of the 2-pt correlation functions. Given that internal covariance estimators mostly underestimate the variance of $\hat{\xi}_\pm$ one can hence conclude that the covariance elements $\mathrm{Cov}_{ij}$ are approximately underestimated by the same amount as the square root of $\mathrm{Cov}_{ii}\mathrm{Cov}_{jj}$ (cf. eqn. \ref{eq:correlation_matrix}).

\begin{figure}
\begin{centering}
\includegraphics[width=0.49\textwidth]{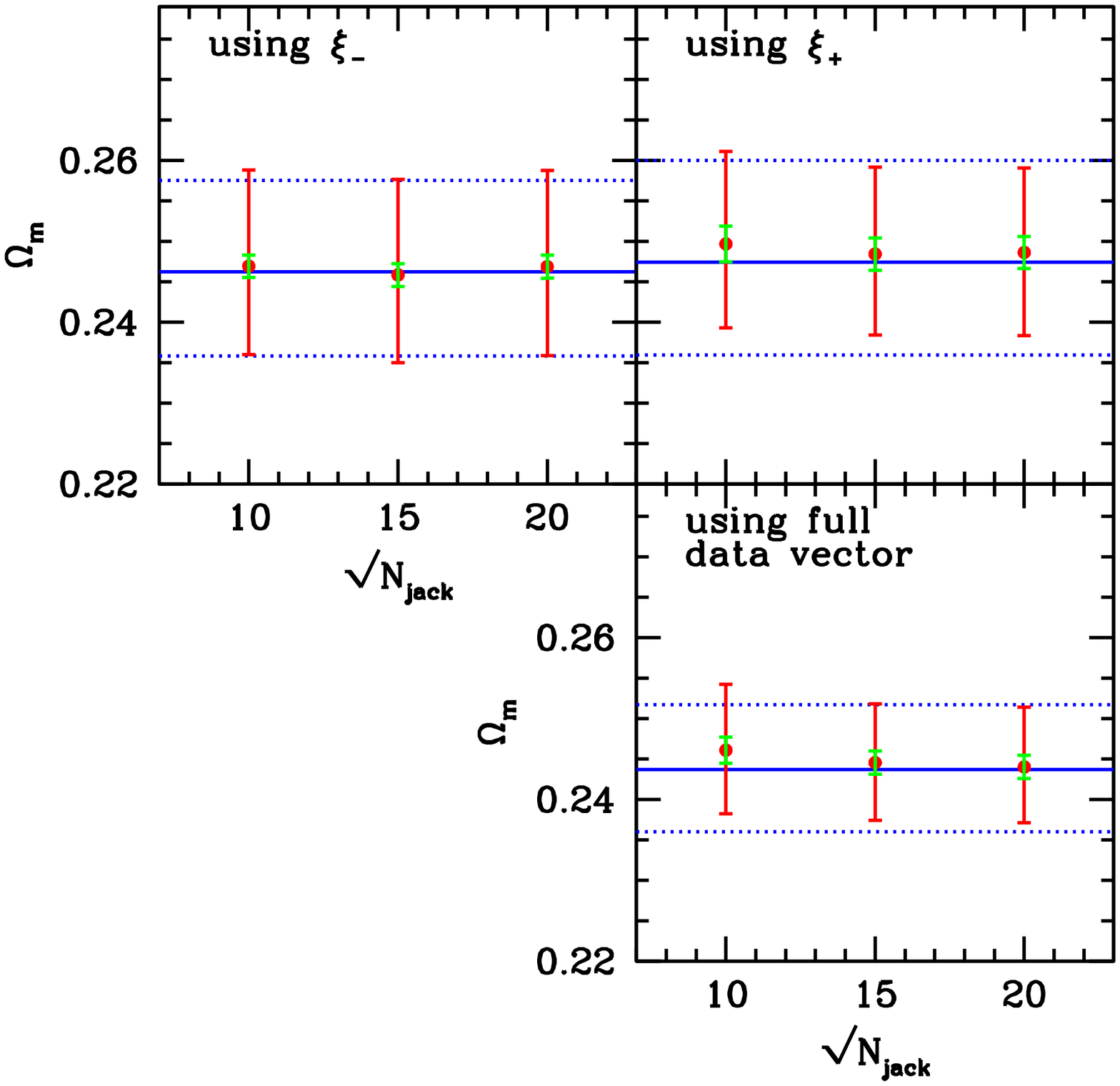}

\includegraphics[width=0.49\textwidth]{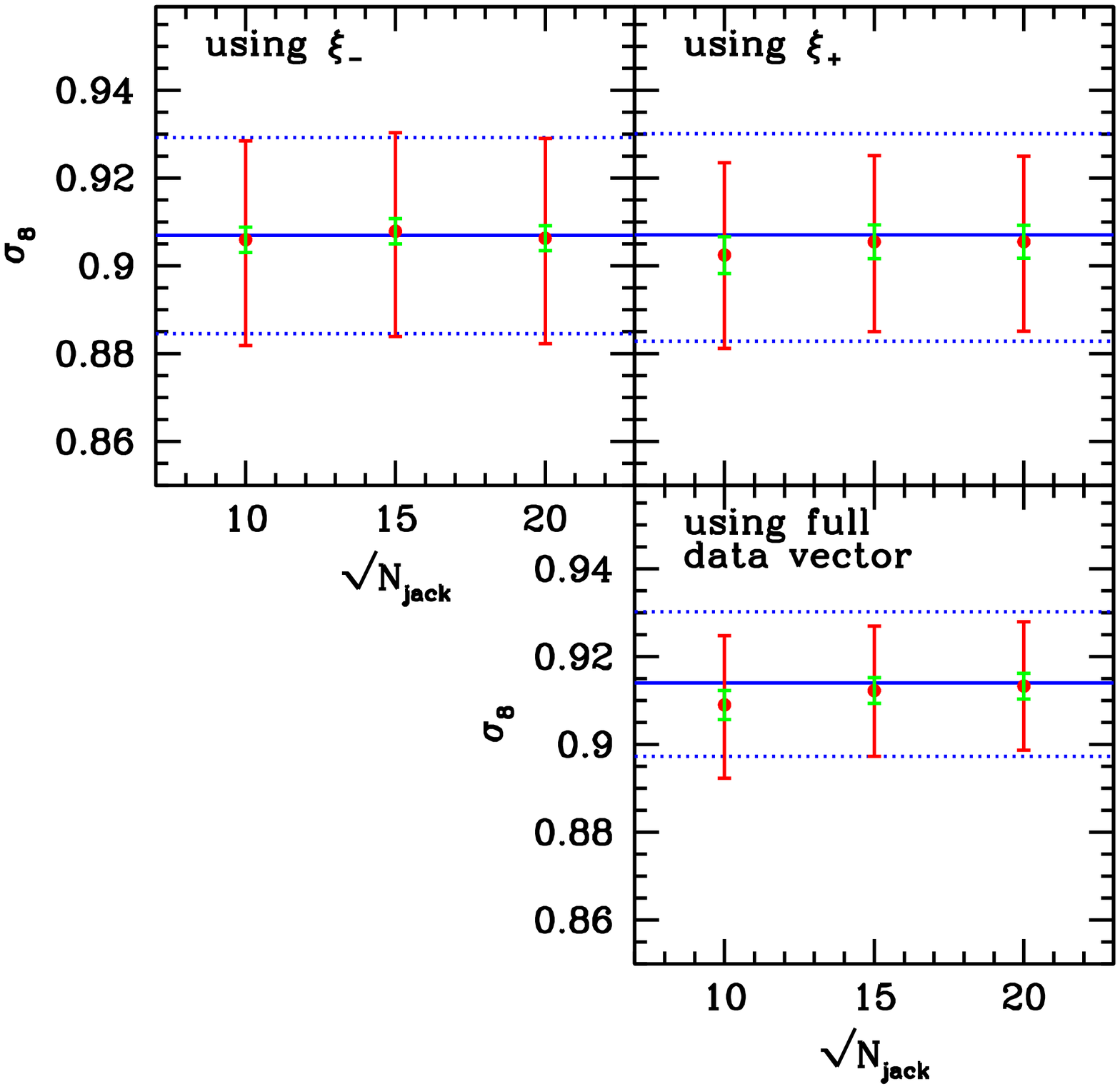}
\end{centering}
 \caption{These plots are identical to the ones shown in figure \ref{fig:constraints}, except that this time the blue lines indicate the mean best fit values and $1-\sigma$ constraints obtained from the empirical covariance matrix, i.e. from the sample covariance of $1000$ independent realizations of setup I.}
 \label{fig:constraints_empirical_covariance}
\end{figure}

In the bottom panel of figure \ref{fig:correlation_matrix} we show the same plot but using the empirical covariance matrix obtained from $1000$ independent realizations of setup I in the lower right half of the plot. The empirical covariance matrix is obviously noisier than the log-normal model covariance matrix. In order to confirm, that there are nevertheless no significant deviations of our simulations from the log-normal input model, we show in figure \ref{fig:constraints_empirical_covariance} again the constraints on $\Omega_m$ and $\sigma_8$ that where presented in figure \ref{fig:constraints}, but this time using the empirical covariance to compute the reference constraints.

\FloatBarrier

\newpage

$\ \nonumber $

\newpage

\section{Likelihood Contours}
\label{sec:appendix}

\begin{figure*}
\begin{centering}
\end{centering}
\includegraphics[width=0.9\textwidth]{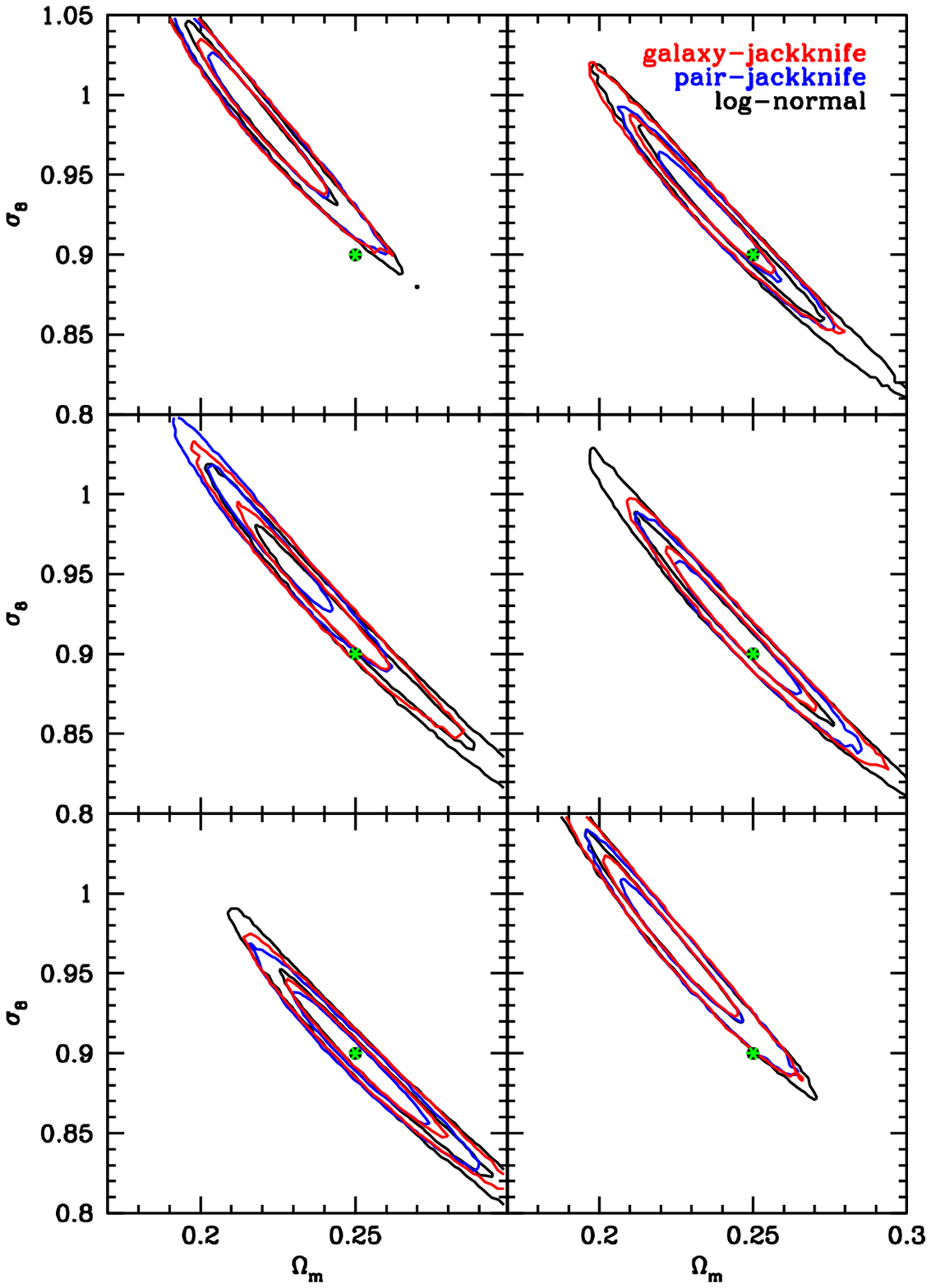}
\caption{$1$- and $2$-$\sigma$ contours in the $\Omega_m$-$\sigma_8$ plane obtained from the first 6 simulations.}
 \label{fig:contours1}
\end{figure*}

\begin{figure*}
\begin{centering}
\end{centering}
\includegraphics[width=0.9\textwidth]{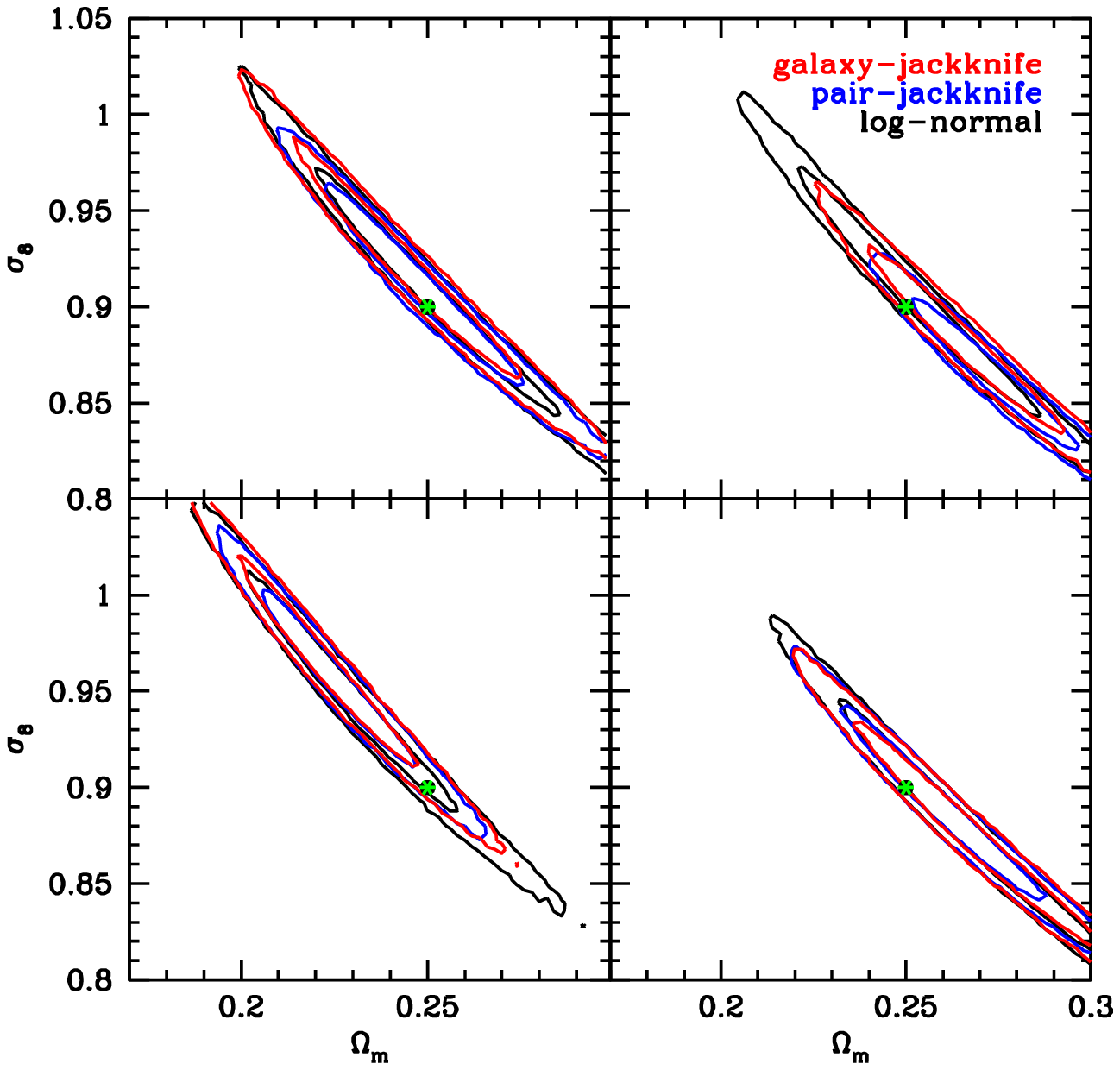}
\caption{$1$- and $2$-$\sigma$ contours in the $\Omega_m$-$\sigma_8$ plane obtained from the remaining 4 simulations.}
 \label{fig:contours2}
\end{figure*}

Figures \ref{fig:contours1} and \ref{fig:contours2} show the $1$- and $2$-$\sigma$ contours in the $\Omega_m$-$\sigma_8$ plane computed with COSMOLIKE when using galaxy-jackknife and pair-jackknife to estimate the covariance matrix (red and blue lines) and compare them to the same contours obtained from the true covariance matrix (black lines). The simulations are configured to mimic the complete, 5 year Dark Energy Survey (cf. section \ref{sec:des_sv} or table \ref{tab:setup}, setup IIb). The only thing that differs from simulation to simulation is the random seed that was used to generate the log-normal fields and the shape noise. The green dots represent the input cosmology of the simulations.

\label{lastpage}

\end{document}